\documentclass[12pt]{article}
\usepackage{amsmath,amssymb,epsfig}
\usepackage{color}
\input{colordvi.tex}
\usepackage[enableskew]{youngtab}
\usepackage{ mathrsfs }

\newcommand{\cb}{{\cal B}}

\newcommand{\cj}{{\cal J}}
\newcommand{\ck}{{\cal K}}
\newcommand{\cn}{{\cal N}}
\newcommand{\cm}{{\cal M}}
\newcommand{\cl}{{\cal L}}

\newcommand{\co}{{\cal O}}

\newcommand{\cs}{{\cal S}}

\newcommand{\cw}{{\cal W}}

\newcommand{\nn}{\nonumber}
\def\eqa{\begin{eqnarray}}
\def\eqae{\end{eqnarray}}
\def\eq{\begin{equation}}
\def\eqe{\end{equation}}
\def\be{\begin{equation}}
\def\ee{\end{equation}}
\def\bea{\begin{eqnarray}}
\def\eea{\end{eqnarray}}
\def\ba{\begin{array}}
\def\ea{\end{array}}
\def\bd{\begin{displaymath}}
\def\ed{\end{displaymath}}

\def\Tr{{\rm Tr}}

\def\>{\rangle}
\def\<{\langle}
\def\a{\alpha}

\def\c{\chi}
\def\del{\delta}
\def\e{\epsilon}

\def\h{\eta}

\def\l{\lambda}

\def\n{\nu}

\def\q{\theta}
\def\r{\rho}

\def\x{\xi}

\def\G{\Gamma}

\def\L{\Lambda}
\def\O{\Omega}  

\def\Q{\Theta}

\def\pa{\partial}
\def\as{asymptotic symmetry}

\def\wt{\widetilde}

\newcommand{\bc}{{\mathbb{C}}}
\newcommand{\br}{{\mathbb{R}}}
\newcommand{\bz}{{\mathbb{Z}}}

\newcommand{\fn}{\footnotemark\footnotetext}

\newcommand{\cpn}{$\bc$P$^{n}$}
\newcommand{\fs}{\lfloor s\rfloor}
\newcommand{\fms}{\lfloor -s\rfloor}
\newcommand{\ft}{\lfloor t\rfloor}
\newcommand{\fmt}{\lfloor -t\rfloor}
\newcommand{\ls}{\lceil s \rceil}
\newcommand{\lt}{\lceil t \rceil}

\makeatletter

\renewcommand\section{\@startsection {section}{1}{\z@}
                                   {-3.5ex \@plus -1ex \@minus -.2ex}
                                   {2.3ex \@plus.2ex}
                                   {\normalfont\large\bfseries}}

\renewcommand\subsection{\@startsection{subsection}{2}{\z@}
                                     {-3.25ex\@plus -1ex \@minus -.2ex}
                                     {1.5ex \@plus .2ex}
                                     {\normalfont\normalsize\bfseries}}

\newcount\hour \newcount\minute
\hour=\time \divide \hour by 60
\minute=\time
\def\now{
\ifnum \hour<13
  \ifnum \hour=0 \advance \hour by 12 \number\hour:\else \number\hour:\fi
     \ifnum \minute<10 0\fi
     \number\minute
\ A.M.
\else \advance \hour by -12 \number\hour:
  \ifnum \minute<10 0\fi
  \number\minute
  \ P.M.
\fi
}

\makeatother

\begin{document}

\baselineskip=18pt
\numberwithin{equation}{section}
\allowdisplaybreaks

\thispagestyle{empty}

\vspace*{-2cm}
\begin{flushright}
\end{flushright}

\begin{flushright}
MCTP-12-08\\
\end{flushright}

\begin{center}

\vspace{2.4cm}

{\bf\Large Symmetries of Holographic Super-Minimal Models}
\\

\vspace*{1.5cm}
{\bf
Kentaro Hanaki and Cheng Peng} \\
\vspace*{0.5cm}

{\it Michigan Center for Theoretical Physics}\\
{\it Department of Physics, University of Michigan}\\
{\it Ann Arbor, MI 48109, USA}

\vspace*{0.5cm}

\end{center}

\vspace{1cm} \centerline{\bf Abstract} \vspace*{0.5cm}

We compute the asymptotic symmetry of the higher-spin supergravity theory in AdS$_{3}$ and obtain an infinite-dimensional non-linear superalgebra, which we call the super-$W_{\infty}[\l]$ algebra. According to the recently proposed supersymmetric duality between higher-spin supergravity in an AdS$_{3}$ background and the 't Hooft limit of the $\cn=2$ \cpn~Kazama-Suzuki model on the boundary, this symmetry algebra should agree with the 't Hooft limit of the chiral algebra of the CFT, $\cs\cw_{n}$. We provide two nontrivial checks of the duality. By comparing the algebras, we explicitly match the lowest-spin commutation relations in the super-$W_{\infty}[\l]$ with the corresponding commutation relations in the 't Hooft limit on the CFT side. We also consider the degenerate representations of the two algebras and find that the spectra of the chiral primary fields are identical.

\newpage
\tableofcontents

\setcounter{page}{1}

\section{Introduction}

The AdS$_{3}$/CFT$_2$ correspondence is an attractive testing-ground for gauge/gravity dualities. On the gravity side, three-dimensional gravity possesses significantly fewer degrees of freedom than higher-dimensional analogues due to the fact that tensor fields with spin greater than one do not have any bulk degrees of freedom, but their dynamics are localized at the boundary. This fact even allows an exact computation of the partition function of the theory for the pure gravity case \cite{mw}. Therefore, gravity on AdS$_{3}$ spacetime is much simpler than its higher-dimensional counterparts. On the field theory side, the Virasoro algebra of the two-dimensional CFT imposes an infinite number of constraints on the dynamics, and this drastically facilitates the analysis of the theory.

Among various versions of the AdS$_{3}$/CFT$_{2}$ duality, the recently proposed duality \cite{gg} between pure gravity coupled to massless higher-spin gauge fields with two massive scalars in an AdS$_{3}$ background and the large-$N$ limit of 2d $\cw_{N}$ minimal models is of great interest. The key ingredients in this conjecture are the higher-spin fields. It has been shown that in a $d$-dimensional background with constant negative curvature, an infinite tower of massless higher-spin fields can be introduced with consistent interactions \cite{va}. Since the proposed CFT dual to this higher-spin theory, the $W_{n}$-minimal model, is in principle solvable at any value of the 't Hooft coupling, this duality is supposed to be easier to study than the previously conjectured duality between higher-spin gravity in AdS$_{4}$ and the 3d $O(N)$ vector model \cite{kp}. Therefore, it serves as a useful tool to understand the connection between the large-$N$ limit of gauge theory and gravity beyond the pure gravity limit \cite{gg}.

Several nontrivial checks have been done on the duality:  including the matching of the symmetries \cite{hr, cfpt, gh,cfp}, the spectra \cite{gg}, the partition functions \cite{gghr} and the correlation functions \cite{pr,akp}. Further studies of spacetime geometry in higher-spin gravity can be found in \cite{clm,gk,agkp, cggr}  and of the higher-spin AdS$_{3}$/CFT$_{2}$ correspondence in \cite{cy,ca,gv,Gaberdiel:2012yb}.

In this paper, we discuss the ${\cal N}=2$ supersymmetric version of the duality \cite{chr}, with a particular emphasis on the correspondence of the symmetries. In the supersymmetric case, it was proposed in \cite{chr} that ${\cal N}=2$ higher-spin supergravity in  AdS$_{3}$ based on the higher-spin algebra $shs[\lambda]$ \cite{bdv1,bdv2}
is dual to the 't Hooft limit of the ${\cal N}=2$ \cpn~minimal model defined in \cite{ks} by the coset
\begin{equation}\label{ksmd}
\frac{\widehat{SU(n+1)_{k}}  \times \widehat{SO(2n)_{1}}}  {\widehat{SU(n)_{k+1}}  \times \widehat{U(1)}_{n(n+1)(k+n+1)}}\;.
\end{equation}
The 't Hooft limit is defined by\footnote{We follow the convention of \cite{bdv1,bdv2} for $shs[\l]$. Their $\l$ is different from $\l$ in \cite{chr} by a factor of two, and that is why there is two in the denominator of the following equation.}
\begin{equation}
n,\; k \to \infty, \quad \lambda = \frac{n}{2(n+k)}\; :\; \text{fixed}\;.
\end{equation}

Although supersymmetry is not necessary to take full control of the theories on both sides, it is still very useful to consider the supersymmetric version of the duality. First, with supersymmetry, calculations are easier thanks to the presence of more symmetry constraints. Secondly, there are new objects we can study such as chiral rings and spectral flow, which provide a larger stage to study the duality. Finally, the higher-spin theory is expected to be related to string theory in the small string tension limit \cite{hr}.  So, to make a connection to superstring theory, it is very natural to consider the supersymmetrized version of the duality.

The paper is organized as follows. In section 2, we review the Chern-Simons formulation of ${\cal N}=2$ higher-spin supergravity based on the higher-spin algebra $shs[\lambda]$ \cite{bdv1,bdv2}. In section 3, we discuss the asymptotic symmetry of the higher-spin supergravity theory and obtain the non-linear super-$W_{\infty}[\lambda]$ algebra.\footnote{This non-linear super-$W_{\infty}[\lambda]$ algebra should be distinguished from the linear super-$W_{\infty}$ algebra obtained in \cite{bdv1,bdv2}. In the rest of the paper, the super-$W_{\infty}[\l]$ algebra means the non-linear version unless otherwise mentioned.} In section 4, we introduce the chiral superalgebra, denoted by $\cs\cw_{n}$, of the dual \cpn~minimal model and provide two non-trivial checks on the correspondence of the symmetries. Finally, we conclude with a discussion in section 5.

As we were in the final stage of the work, the paper \cite{superW} appeared with some overlapping results.

\section{Higher-spin supergravity as a Chern-Simons Theory}

In \cite{at, w88}, it was shown that classical three-dimensional Einstein gravity in an AdS$_3$ background can be reformulated as an $SL(2,{\br})_L \times SL(2,{\br})_R$ Chern-Simons theory. Define the $SL(2,{\br})_L \times SL(2,{\br})_R$ connections
\begin{equation}
A = (\omega^a  + \frac{1}{\ell} e^a) J^a\;, \qquad
\tilde{A} = (\omega^a  - \frac{1}{\ell} e^a )\tilde{J}^a\;, \qquad a=1,2, 3\,,
\end{equation}
where $\ell$ is the radius of AdS$_3$, $J^a$ are generators of $SL(2,{\br})_L$, and $\tilde{J}^a$ are generators of $SL(2,{\br})_R$. The Einstein-Hilbert action can then be written as
\begin{equation}
I_{EH} = I_{CS}(A) - I_{CS}(\tilde{A})\;, \qquad
I_{CS} (A)= \frac{k_{CS}}{4 \pi}  \int_{\cm} \mbox{Tr}(A \wedge d A + \frac{2}{3} A \wedge A \wedge A)\;,
\end{equation}
where the Chern-Simons level $k_{CS}$ is related to the Newton's constant in AdS$_{3}$ spacetime as
\begin{equation}
k_{CS}=\frac{\ell}{4G_3}\,,
\end{equation}
and the trace Tr is taken over gauge indices throughout the paper. One can show that $e^a$ and $\omega^a$ behave in the same way as the vielbein and spin connection, respectively, in Einstein gravity on-shell \cite{w88}. This formulation is extended to particular types of higher-spin theories with and without supersymmetry in \cite{Blencowe:1988gj}. In this section, we discuss how to extend this Chern-Simons formulation to ${\cal N}=2$ higher-spin supergravity based on $shs[\lambda]$.

\subsection{Supersymmetric higher-spin algebra $shs[\lambda]$}

The ${\cal N}=2$ higher-spin supergravity theory is formulated as a Chern-Simons theory based on the super-higher-spin algebra $shs[\lambda]$. We start with a briefly review of this algebra.

$shs[\lambda]$ is a one-parameter family of Lie superalgebras \cite{bdv1,bdv2}. It admits $\cn=2 $ supersymmetry and consists of two sets of bosonic generators $L^{(s)\pm}_m$ as well as two sets of fermionic generators $G^{(s)\pm}_r$. The integer $s$ satisfies $s \ge 2$ for $L^{(s)+}_m$ and $G_r^{(s)\pm}$, and $s \ge 1$ for $L^{(s)-}_m$. The integer $m$ satisfies $|m| < s$ and $r$ is a half-integer satisfying $|r|<s-1$. The algebraic structure of $shs[\lambda]$ is provided in Appendix \ref{shsapp}. Here, we only address two points \cite{bdv2}
\begin{itemize}
\item The $shs[\lambda]$ algebra contains an $Osp(1,2)$ algebra generated by $L^{(2)+}_m$ and $G^{(2)+}_r$ as a subalgebra. $(L^{(s)+}_{m}, G^{(s)+}_{r})$ and $(L^{(s)-}_{m}, G^{(s+1)-}_{r})$ form an $\cn=1$ supermultiplet of the $Osp(1,2)$ subalgebra with $SL(2)$ spins $(s, s-1/2)$ and $(s, s+1/2)$, respectively. The  $Osp(1,2)$ generators $L^{(2)+}_m$ and $G^{(2)+}_r$, together with $L^{(1)-}_m$ and $G^{(2)-}_r$ generate an $Osp(2,2)$ subalgebra, where $L^{(1)-}_{0}$ corresponds to the $R$-charge of the superalgebra.
\item The $shs[\lambda]$ algebra can be truncated at a special value of $\lambda$. For $\lambda = 1/4$, the $+$ sector (of generators with a ``$+$" index) and $-$ sector decouple, and the $+$ sector reduces to the ${\cal N}=1$ superalgebra, which was used to construct ${\cal N}=1$ higher-spin supergravity in \cite{Fradkin:1987ah,Blencowe:1988gj}.
\end{itemize}
In the following, we relabel the generators in $+$ sector and $-$ sector as
\begin{equation}
L^{(s)}_m = L^{(s)+}_m\,, \qquad L^{(s+1/2)}_m = L^{(s)-}_m\,, \qquad
G^{(s)}_r = G^{(s)+}_r\,, \qquad G^{(s+1/2)}_r = G^{(s+1)-}_r\,,
\end{equation}
for notational simplicity.

The ${\cal N}=2$ higher-spin supergravity theory is formulated as a Chern-Simons theory based on the gauge group $shs[\lambda]_L \times shs[\lambda]_R$.\footnote{There can be several ways to embed the gravity sector into the higher-spin algebra. We take $L^{(2)}_m$, $L^{(3/2)}_m$, $G^{(2)}_r$ and $G^{(3/2)}_r$ as the generators associated with the ${\cal N}=2$ supergravity.} The $shs[\lambda]_L \times shs[\lambda]_R$ super-connections are given by
\begin{equation}
\Gamma = {\displaystyle \sum_{s,m}} A^{(s)}_{m} L^{(s)}_{m} + {\displaystyle \sum_{s,r}}\psi^{(s)}_{r} G^{(s)}_{r}\;, \qquad
\tilde{\Gamma} ={\displaystyle \sum_{s,m}} \tilde{A}^{(s)}_{m} \tilde{L}^{(s)}_{m} + {\displaystyle \sum_{s,r}} \tilde{\psi}^{(s)}_{r} \tilde{G}^{(s)}_{r}\;,
\end{equation}
where $A$ and $\tilde{A}$ are expressed using (particular linear combinations of higher-spin analogues of) vielbeins and spin connections as
\begin{equation}
A^{(s)}_m = \omega^{(s)}_m + \frac{1}{\ell} e^{(s)}_m\;, \qquad\qquad
\tilde{A}^{(s)} = \omega^{(s)}_m - \frac{1}{\ell} e^{(s)}_m\,,
\end{equation}
where $\ell$ is the AdS radius and the action is obtained as a difference of two Chern-Simons actions
\begin{equation}
I_{SHS} = I_{CS}(\Gamma) - I_{CS}(\tilde{\Gamma})\;.
\end{equation}
With the help of the equations of motion, $e^{(2)}$ and $\omega^{(2)}$ are identified with the vielbein and spin connection and $\psi^{(2)}$, $\tilde{\psi}^{(2)}$ are identified as two sets of gravitinos.

\subsection{Boundary conditions, constraints and gauge fixing}

Now that the action is obtained, we discuss how one defines a consistent theory based on this action. First of all, in order for the variational principle to be well-defined, the variation of the action should not depend on the variation of the field at the boundary. The variation of the action is
\begin{equation}
\delta I_{CS} = -\frac{k_{CS}}{4 \pi} \int_{\partial \cm} \mbox{Tr} (\Gamma_+ \delta \Gamma_- - \Gamma_- \delta \Gamma_+ ) -\frac{k_{CS}}{4 \pi} \int_{\cm}(\mbox{e.o.m.})\;,
\end{equation}
where we use coordinates $(t,\r,\q)$ to parameterize the spacetime manifold $\cm$,  and define $x^{\pm}=t/\ell \pm \theta$. One can set the boundary contribution to zero by
fixing $\G_{-}$ on the boundary:
\begin{equation}
\Gamma_-\big|_{\partial \cm} = 0\;.
\label{minimalBC}
\end{equation}
We call this ``minimal'' boundary condition to distinguish it from the boundary condition we impose in the next section from which we obtain the asymptotic algebra super-$W_\infty[\lambda]$.

We also need to fix the gauge degrees of freedom. We choose the gauge fixing conditions, following \cite{cfpt}, as
\begin{equation}
\Gamma_\rho = b^{-1}(\rho) \partial_{\rho} b(\rho)\;,
\label{minimalGF}
\end{equation}
where $b(\rho)$ is an arbitrary, but fixed, function of the radial coordinate $\rho$ and takes values in the group $shs[\l]$. In the later sections, we take $b(\rho)$ to be
\begin{equation}\label{bl0}
b(\rho) = e^{\rho L^{(2)}_0}
\end{equation}
for $shs[\lambda]$ Chern-Simons theory, but the discussions in this section are independent of the choice of $b(\rho)$. In the action, there is no time derivative of $A_t$, implying that there is a constraint. The variation of the action with respect to $A_t$ yields
\begin{equation}\label{gq}
\partial_\rho \Gamma_{\theta} + [\Gamma_\rho, \Gamma_\theta] = 0\;.
\end{equation}
This can be solved uniquely by
\begin{equation}\label{gmfx}
\Gamma_\theta = b^{-1}(\rho) \gamma(t, \theta ) b(\rho)\;,
\end{equation}
where $\gamma(t, \theta)$ is an arbitrary function of $t$ and $\theta$. Therefore, the degrees of freedom are reduced to $\gamma(t,\theta)$ by the gauge fixing and constraints.

\subsection{Global symmetry}

We are ready to discuss the global symmetry of the theory with our minimal boundary condition \eqref{minimalBC}. The global symmetry is defined to be the residual symmetry after the gauge fixing that leaves the boundary condition \eqref{minimalBC} and the gauge fixing condition \eqref{minimalGF} invariant. The invariance of the gauge fixing condition \eqref{minimalGF}, $\delta \Gamma_\rho = 0$, implies that the gauge transformation parameter $\Lambda$ should satisfy
\begin{equation}
\partial_{\rho} \Lambda + [\Gamma_{\rho}, \Lambda] = 0 \;.
\end{equation}
This can be solved uniquely by
\begin{equation}\label{lmfx}
\Lambda = b^{-1}(\r) \lambda(t,\theta) b(\r)\;.
\end{equation}
Now, the invariance of the boundary condition \eqref{minimalBC} imposes a further constraint
\begin{equation}
\partial_- \Lambda\big|_{\partial \cm} = 0\;.
\end{equation}
This implies that
\begin{equation}\label{resgaupar}
\lambda(t,\theta) = \lambda(x^+)\;.
\end{equation}
So, the time dependence of the transformations are fixed by the $\theta$ dependence. Thus the gauge degrees of freedom are completely fixed to $\lambda(x^+)$.

What we are interested in is to discuss the global algebra generated by these transformations. Note that the global charge of this algebra is given by \cite{rt, Benguria:1976in}
\begin{equation}
Q[\Lambda] = - \frac{k_{CS}}{2 \pi} \int_{\partial \Sigma} d \theta~ \mbox{Tr} (\Lambda \Gamma_\theta)\;,
\end{equation}
where $\Sigma$ is a constant time slice. From our boundary conditions and gauge fixing, we have \eqref{gmfx} and \eqref{lmfx}, which lead to
\begin{equation}
Q[\Lambda] = - \frac{k_{CS}}{2 \pi} \int_{\partial \Sigma} d\q~ \mbox{Tr}(\lambda(\theta) \gamma(\theta))\;.
\label{chargeIntegral}
\end{equation}
From this expression, given that the symmetry transformation parameters are $\lambda(\q)$, the generators of the global symmetries are $\gamma(\theta)$ and their algebra is obtained by considering the symmetry transformations of $\gamma(\theta)$. Namely,
\begin{equation}
\delta \gamma(\theta) = \{Q, \gamma(\theta)\} =  - \frac{k_{CS}}{2 \pi} \int d\q'~ \lambda(\theta') \{\gamma(\theta'),\gamma(\theta)\}\;,
\label{Transformation}
\end{equation}
where $\{\cdot,\cdot\}$ is the Poisson bracket. $\delta \gamma(\theta)$ on the left hand side can be derived from the original gauge transformation of $\Gamma_\theta$. To see this, note that the gauge transformation of $\Gamma_\theta$ is given by
\begin{eqnarray}
\delta \Gamma_{\theta} &=& \partial_{\theta} \Lambda - [\Gamma_{\theta}, \Lambda]\nonumber\\
&=& b^{-1}(\r) (\partial_{\theta} \lambda(\theta) - [\gamma(\theta), \lambda(\theta)])b(\r)\;.
\end{eqnarray}
Then, by comparing this with $\delta \Gamma_{\theta} = b^{-1} \delta \gamma(\theta) b$, one obtaines
\begin{eqnarray}
\delta \gamma(\theta) = \partial_\theta \lambda(\theta) - [\gamma(\theta), \lambda(\theta)]\;.
\label{gammaTransformation}
\end{eqnarray}
If one expands $\gamma(\theta)$ in terms of the generators of the gauge group $T^a$ as $\gamma(\theta) = \sum \gamma^a(\theta) T^a$, then the transformations \eqref{gammaTransformation} can be reproduced by the following Poisson bracket:
\begin{equation}\label{Poissongg}
\{\gamma^a(\theta), \gamma^b(\theta')\} = \frac{2 \pi}{k_{CS}} \left[K^{ab} \delta'(\theta - \theta') - f^{ab}{}_c \gamma^c(\theta) \delta(\theta - \theta') \right]\;,
\end{equation}
where $K^{ab}$ is the inverse of the Killing form $K_{ab}$ and $f^{ab}{}_c$ are the structure constants of the gauge group. One can expand $\gamma^a(\theta)$ in modes as
\begin{equation}
\gamma^a(\theta) = \frac{1}{k_{CS}} \sum\limits_{n=-\infty}^{\infty} \gamma^a_m e^{- i m \theta}\;.
\end{equation}
Then, we get the affine Kac-Moody algebra associated with the gauge group:
\begin{equation}
\{ \gamma^a_m, \gamma^b_n \} = imk_{CS} K^{ab} \delta_{m+n,0} - f^{ab}{}_c \gamma_{m+n}^c\;.
\end{equation}
Note that this result is true for general gauge group. As a summary of this section, we reviewed how higher-spin supergravity is realized as a Chern-Simons theory and the global symmetry of the theory with the minimal boundary condition \eqref{minimalBC}. In the next section, we impose a more restrictive boundary condition and see the super-$W_{\infty}[\lambda]$ is realized as the asymptotic symmetry.

\section{Super-$W_{\infty}[\l]$ algebra as the asymptotic symmetry}

The goal of this section is to obtain the non-linear super-$W_\infty[\lambda]$ as the asymptotic symmetry by imposing additional boundary conditions. Super-$W_\infty[\lambda]$ is a higher-spin extension of the ${\cal N}=2$ super-Virasoro algebra. The boundary condition to obtain the super-Virasoro algebra from the affine Kac-Moody algebra is known in the literature \cite{b99,hms}, and we use the same boundary condition and extend their analysis to higher spin cases.

\subsection{Boundary condition for super-$W_{\infty}[\lambda]$ algebra}

In order to obtain the super-$W_{\infty}[\lambda]$ symmetry, we impose a boundary condition \footnote{This boundary condition \eqref{WBoundaryConditions} has been extensively studied in three-dimensional gravity and its supersymmetric extensions \cite{bh,b95,b98,b99,hms}.}
\begin{equation}
( \Gamma - \Gamma_{AdS_3})\big|_{\partial \cm} = {\cal O}(1)\;,
\label{WBoundaryConditions}
\end{equation}
in addition to the minimal boundary condition \eqref{minimalBC}, where $\Gamma_{AdS}$ is the gauge field configuration corresponding to the global $AdS$ geometry and is given explicitly by
\begin{equation}
\Gamma_{AdS_3} = \left[ e^{\rho} L_1^{(2)} + \frac{1}{4} e^{-\rho} L_{-1}^{(2)} \right] d\theta + b(\r)^{-1} \partial_\rho b(\r) d \rho + \G_t dt\;.
\label{gaugeAdS}
\end{equation}
where $b(\r)$ is the same as that in \eqref{bl0}. The boundary condition \eqref{WBoundaryConditions} imposes constraints on $\gamma(\theta)$. To see that, we expand $\gamma(\theta)$ in terms of the $shs[\lambda]$ generators as
\begin{equation}\label{oldgm}
\gamma(\theta) = {\displaystyle \sum_{s,m}} a^{(s)}_m(\theta) L^{(s)}_m + {\displaystyle \sum_{s,r}}\psi^{(s)}_r(\theta) G^{(s)}_r\;.
\end{equation}
This, together with \eqref{minimalGF} and \eqref{gmfx}, fixes the super-connection as
\begin{equation}\label{temp1}
\Gamma = b^{-1}(\r) \left(a^{(s)}_m L^{(s)}_m + \psi^{(s)}_r G^{(s)}_r \right) b(\r) d \q + b^{-1}(\r)  \partial_\rho b(\r)  d \rho + \Gamma_t dt\;,
\end{equation}
where the repeated indices are summed over. Here, the $\Gamma_t$ is equal to $\Gamma_\theta$ at the boundary due to the boundary condition \eqref{minimalBC}, though in the bulk, there is no restriction on it. This time component, however, is expected not to affect the global symmetry of the theory because the charge integral \eqref{chargeIntegral} is taken on a constant time slice, and the dependence on the time component of the gauge field disappear. Therefore, we will not discuss $\Gamma_t$ in the rest of the paper.

The $shs[\l]$ commutation relations read
\begin{equation}
[L_0^{(2)}, L^{(s)}_m] = - m L^{(s)}_m, \qquad [L_0, G^{(s)}_r] = - r G^{(s)}_r\;.
\end{equation}
which reflect the fact that the commutator of any generator with $L^{(2)}_0$ just gives the conformal weight of the generator.  Together with the Baker-Campbell-Hausdorff formula and \eqref{bl0}, \eqref{temp1} can be rewritten as
\begin{equation}\label{gmlt}
\Gamma = \left( e^{ m \rho} a^{(s)}_m L^{(s)}_m + e^{ \rho r } \psi^{(s)}_r G^{(s)}_r\right) d \theta + b^{-1}(\r) \partial_{\rho} b(\r) d \rho + \Gamma_t dt\;.
\end{equation}
The boundary condition \eqref{WBoundaryConditions} implies that, at the boundary $\r\to\infty$, the difference between \eqref{gmlt} and \eqref{gaugeAdS} is order one. This imposes the following constraints:
\begin{eqnarray}
a^{(2)}_1 &=& 1\;,\\
 \qquad a^{(s)}_m &=& 0 \qquad (s \ge 3, \; m > 0)\;, \qquad \psi^{(s)}_r= 0 \quad ( i > 0 )\;.\label{1stcst}
\end{eqnarray}
The constraints \eqref{1stcst} are first class because the Poisson bracket, given in \eqref{Poissongg}, between any pair of them closes into a linear combination of \eqref{1stcst}.\footnote{Note that a Poisson bracket between positive frequency modes close into a linear combination of positive frequency modes.} Therefore, each of these first class constraints generates a gauge symmetry. These $(\lfloor s\rfloor -1)+(\lfloor s\rfloor -1)$ gauge symmetries are fixed by the following $(\lfloor s\rfloor -1)+(\lfloor s\rfloor -1)$ gauge fixing conditions
\begin{equation}\label{2ndcst}
a^{(s)}_m = 0 \quad ( -\fs+1 < m \le 0), ~~\qquad \psi^{(s)}_r = 0 \quad ( \fms + 3/2 < m < 0)\;,
\end{equation}
where $\lfloor\cdot\rfloor$ is the ``floor'' function. These conditions are second class because generally commutators $[a^{(s)}_m, a^{(s)}_{-m+1}]$ and $\{\psi^{(s)}_r, \psi^{(s)}_{-r+1}\}$ close into certain linear combinations of constraints plus $a^{(2)}_1$, which is non-vanishing under the constraints. Therefore, the only unconstrained fields are
\begin{equation}
a^{(s)}_{1-\fs}\equiv  \frac{2 \pi}{k_{CS} N^B_s}a_{s}\;,\qquad \qquad \psi^{(s)}_{3/2+\fms}\equiv  \frac{2 \pi}{k_{CS} N^F_s}\psi_{s}\;.
\end{equation}
where the normalization functions are defined by \mbox{$N^B_s = \mbox{Tr} (L^{(s)}_{-\fs+1} L^{(s)}_{\fs-1})$} and
${N}^F_s =$ \\$\text{Tr} ( G^{(s)}_{\lceil s\rceil-3/2}G^{(s)}_{\fms+3/2})$ with $\lceil \cdot\rceil$ being the ``ceiling'' function. Their values at small $s$ are listed in Appendix \ref{shsapp}.

The $\gamma(\q)$ in \eqref{oldgm} is thus constrained by \eqref{1stcst}, \eqref{2ndcst} to be
\begin{equation}
\gamma(\theta) = L_{1} + \frac{2 \pi}{k_{CS}}{\displaystyle \sum_{s \ge 3/2, s \in \frac{1}{2} {\bz}}} \left( \frac{1}{N^B_s} a_s(\theta) L^{(s)}_{-\fs+1} + \frac{1}{{N}^F_s} \psi_s(\theta)G^{(s)}_{\fms + 3/2} \right)\;.
\label{fixedFields}
\end{equation}
This is the most general form of the super-connection that is compatible with the boundary condition \eqref{WBoundaryConditions}. In the next subsection, we will derive the symmetry algebra that leaves the form of the super-connection \eqref{fixedFields} invariant.

\subsection{Super-$W_{\infty}[\l]$ symmetries}

We are now ready to discuss the asymptotic symmetry under the boundary condition \eqref{WBoundaryConditions}. For convenience, we expand the gauge transformation parameter $\Lambda$, and the gauge variations of fields $a(\theta)$ and $\psi(\theta)$ in terms of the generators of $shs[\lambda]$ as
\begin{equation}
\Lambda = {\displaystyle \sum_{s \ge 3/2, s \in \frac{1}{2}{\bz}}} \left( {\displaystyle \sum_{m \in {\bz}}} \x_m^{(s)} L_m^{(s)} + {\displaystyle \sum_{r \in {\bz} + 1/2}} \epsilon_r^{(s)} G_r^{(s)} \right)\;,
\end{equation}
\begin{equation}\label{apvari}
 \delta a = \sum_{s} \sum_m c^B_{s,m} L^{(s)}_m\;, \qquad \delta \psi = \sum_s \sum_r c^F_{s,r} G^{(s)}_r \;,
\end{equation}
where we omit the argument $\theta$. Then, under the gauge transformation \eqref{gammaTransformation}, $c^B_{s, m}$ and $c^F_{s, r}$ are found to be
\begin{eqnarray}
c^B_{s,m} &=& \partial_+ \x^{(s)}_m + (- m +\fs) \x_{m-1}^{(s)} \nonumber\\
& &+{\displaystyle \sum_t}  \left[{\displaystyle \sum_u} a^{(t)}_{-\ft+1} ~\x_{m + \ft - 1}^{(s+u-t)}~ g^{t, s+u-t}_u\big(-\ft+1,m+\ft-1;\lambda\big)\right.\nonumber\\
& &\left.-{\displaystyle \sum_v}\psi^{(t)}_{\fmt +3/2} ~\epsilon^{(s+v-t)}_{m+\fmt-3/2}~
\tilde{g}_v^{t,s+v-t}\big(\fmt +3/2,m+\fmt-3/2;\lambda\big) \right]\;,\label{cbsm}\\
[3mm]
c^F_{s,r} &=& \partial_+ \epsilon^{(s)}_r + (-r + \lfloor s+1/2 \rfloor - 1/2) \nonumber\\
& &+ {\displaystyle \sum_t} \left[{\displaystyle \sum_v} a^{(t)}_{-\ft+1} ~\x_{r + \ft - 1}^{(s+v-t)} ~h^{t, s+v-t}_v\big(-\ft+1,r+\ft-1;\lambda\big)\right.\nonumber\\
& &\left.-{\displaystyle \sum_u} \psi^{(t)}_{\fmt +3/2} ~\epsilon^{(s+u-t)}_{r+\fmt-3/2}
~\tilde{h}_u^{t,s+u-t}\big(\fmt+3/2,r+\fmt-3/2;\lambda\big) \right]\;,\label{cfsr}
\end{eqnarray}
where $g^{st}_{u}$, $\tilde{g}^{st}_u$, $h^{st}_u$ and $\tilde{h}^{st}_u$ are the structure constants of $shs[\l]$ and we provide examples of their explicit expressions in Appendix \ref{shsapp}. The ranges of summations in \eqref{cbsm} are
\begin{subequations}\label{rangeb}
\begin{align}
\text{max}(1+|m+\lt-1|,1+\fs-\ft)\leq \lfloor s+u-t \rfloor  \qquad &\text{and} \qquad 1\leq u \leq 2s-\frac{1}{2}\\
\text{max}(\frac{3}{2}+\big| m+\ft-\frac{3}{2} \big|,2+\ls-\ft)\leq \lceil s+v-t \rceil  \qquad &\text{and} \qquad 1\leq v \leq 2s-\frac{1}{2}
\end{align}
\end{subequations}
The ranges of summations in \eqref{cfsr} are
\begin{subequations}\label{rangef}
\begin{align}
\text{max}(\frac{3}{2}+| m+\ft-1 |,1+\ls-\ft)\leq \lceil s+v-t \rceil  \qquad &\text{and} \qquad 1\leq v \leq 2s-\frac{1}{2}\\
\text{max}(1+|m+\lt-1|,1+\ls-\lt)\leq \lfloor s+u-t \rfloor  \qquad &\text{and} \qquad 1\leq u \leq 2s-\frac{1}{2}
\end{align}
\end{subequations}
The global symmetry consists of the transformations which preserve the structure \eqref{fixedFields}. In terms of $c^B_{s,m}$ and $c^F_{s,r}$, preserving \eqref{fixedFields} implies:
\begin{equation}\label{ccon}
c_{s,m}^B = 0 \quad \mbox{for }m \neq - \fs+1\; ~~\text{ and }\quad c_{s,r}^F = 0 \quad \mbox{for }r \neq \fms+3/2\;.
\end{equation}
One can solve these conditions. As a result, we find that the only independent transformation parameters are
\begin{equation}
\eta_s \equiv \x_{\fs-1}^{(s)} \qquad\text{and}\qquad \epsilon_s \equiv \epsilon_{
\lceil s\rceil-3/2}^{(s)}
\end{equation}
and all other parameters can be expressed in terms of these independent parameters.

Once all the transformation parameters $\x_{m}^{(s)},~\e_{r}^{(s)}$ are solved in terms of $\h_{s}$ and $\e_{s}$, one can compute the variation of $a$'s and $\psi$'s \eqref{apvari}. These variations can be written as:
\begin{eqnarray}
\nn \delta_s^B a_t&=& \frac{k_{CS}}{2 \pi} N^B_t c^B_{t,1-\ft}(\eta_s)\;, ~\qquad
\delta_s^F a_t= \frac{k_{CS}}{2 \pi} N^B_t c^B_{t,1-\ft}(\e_s)\;,\\
\delta_s^B \psi_t &=& \frac{k_{CS}}{2 \pi} {N}^F_t c^F_{t,\fmt+3/2} (\eta_s)\;,
\quad
\delta_s^F \psi_t = \frac{k_{CS}}{2 \pi} {N}^F_t c^F_{t,\fmt+3/2} (\epsilon_s)\;.\label{vat}\end{eqnarray}
where $\delta_s^{B(F)}$ represents a variation corresponding to the bosonic (fermionic) generator with spin $s$. The argument $(\h_{s})$ means that we turn on $\h_{s}$ and set $\e_{s}$ to zero, and similar for $(\e_{s})$.

Calculating the global symmetry algebra amounts to solve \eqref{ccon} and express all $\x_m^{(s)}$ and $\e^{(s)}_r$ in \eqref{vat} in terms of $\h_s$ and $\e_s$. While solving \eqref{ccon} in full generality is a difficult task, we focus on the variations including the lower spin generators. First of all, we find that the variations including $s =3/2$ and $s=2$ are given by
\begin{subequations}\label{supervirasoro}
\begin{align}
\delta^B_2 a_2 &= 2 a_2 \eta' + a_2' \eta - \frac{k_{CS}}{4 \pi} \eta'''\label{Virasoro}\;,\\
\delta^B_2 a_{3/2} &= 0\;,\\
\delta^B_{3/2} a_{3/2} &= -\frac{k_{CS}}{\pi} \eta'\label{adsjj}\;,\\
\delta^B_2 \psi_2 &=\frac{3}{2} \psi_2 \eta' + \psi_2' \eta + \frac{\pi}{k_{CS}} a_{3/2} \psi_{3/2} \eta\;,\\
\delta^B_2 \psi_{3/2} &= \frac{3}{2} \psi_{3/2} \eta' + \psi_{3/2}' \eta + \frac{\pi}{k_{CS}} a_{3/2} \psi_2 \eta\;,\\
\delta^B_{3/2} \psi_2 &= \psi_{3/2} \eta\;,\\
\delta^B_{3/2} \psi_{3/2} &= \psi_2 \eta\;,\\
\delta^F_2 \psi_2 &= -2 a_2 \epsilon + \frac{\pi}{k_{CS}} a_{3/2}^2 \epsilon + \frac{k_{CS}}{\pi} \epsilon''\;,\\
\delta^F_2 \psi_{3/2} &= 2 a_{3/2} \epsilon' + a_{3/2}' \epsilon\;,\\
\delta^F_{3/2} \psi_{3/2} &= 2 a_2 \epsilon - \frac{\pi}{k_{CS}} a_{3/2}^2 \epsilon - \frac{k_{CS}}{\pi} \epsilon''\;,
\end{align}
\end{subequations}
where $\h'$ represents $\frac{\pa \h(\q)}{\pa\q}$ and the subscripts $s$ of $\h_{s}$, which are the same $s$ as in the $\delta_s^{B(F)}$ , are omitted.

To reproduce the standard form of the ${\cal N}=2$ super-Virasoro algebra, one first needs to, as in \cite{hms}, redefine the $a_{2}$ as
\begin{equation}\label{modify}
a_2^{\text{SVA}} = a_2 + \frac{\pi}{2 k_{CS}} a_{3/2} a_{3/2}
\end{equation}
and the fermionic generators as $\psi^{\text{SVA}}_+ = \frac{1}{2}(\psi_2 + \psi_{3/2})$ and $\psi^{\text{SVA}}_- = \frac{1}{2}(\psi_2 - \psi_{3/2})$, where $\psi_{\pm}^{\text{SVA}}$ has $U(1)_R$ charge $\pm 1$. Then, one can convert the variation into Poisson bracket using \eqref{Transformation} and expand the fields into modes using
\begin{equation}\label{adsmode}
\co(\q)=\frac{1}{2\pi} \sum\limits_{p\in \bz} \co_{p} e^{i p \q}\,.
\end{equation}
Plugging this into the Poisson bracket gives the commutators between the modes. Finally, one needs to modify the zero mode of $a_{2}^{\text{SVA}}$ as
\begin{equation}\label{modeshift}
a_{2,p}^{\text{SVA}} \to a_{2,p}^{\text{SVA}} - \frac{k_{CS}}{4} \del_{p,0}\,.
\end{equation}
For example, the commutator between two $a_2^{\text{SVA}}$'s reproduces the Virasoro algebra:
\begin{equation}\label{centralterm}
[(a_2^{\text{SVA}})_{m},~(a_2^{\text{SVA}})_{n}]=(m-n)(a_2^{\text{SVA}})_{m+n}+\frac{c_\text{AdS}}{12}(m^{3}-m)\del_{m+n,0},\quad \text{where}~~ c_{\text{AdS}}=6 k_{CS}\,.
\end{equation}
This is how we obtain the standard form of the ${\cal N}=2$ super-Virasoro algebra.

We have also shown that the variations of $a_s$ and $\psi_s$ with $s=3/2,2$ with respect to generators with the spin greater than two satisfy
\begin{subequations}\label{hsvir}
\begin{align}
\delta_n^B a_2 &= \lfloor n \rfloor a_n \eta' + ( \lfloor n \rfloor -1) a_n' \eta - \frac{k_{CS}}{4 \pi} \eta''' \delta_{n,2}\label{BBhigher}\;,\\
\delta_n^F a_2 &= (\lfloor -n \rfloor+1/2) \psi_n \epsilon' + ( \lfloor -n \rfloor+3/2) \psi_n' \epsilon + FB_{n,2}\label{FBhigher}\;,\\
\delta_n^B a_{3/2} &= 0\;,\qquad\qquad \delta_{n-1/2}^B a_{3/2} = 0\,,\label{JBhigher}\\
\delta_{n-1/2}^F a_{3/2} &= -\psi_n \epsilon\;,\\
\delta_n^F a_{3/2} &= -\psi_{n-1/2} \epsilon\;,\\
\delta_n^B \psi_2 &= (n-1/2)\psi_{n} \eta' + (n-1) \psi_{n}' \eta + BF_{n,2}\;,\\
\delta_{n-1/2}^B \psi_{2} &= \psi_{n-1/2} \eta\;,\\
\delta_n^F \psi_{2} &= -2 a_n \epsilon + FF_{n,2}\;,\\
\delta_{n-1/2}^F \psi_2 &= (2-2n) a_{n-1/2} \epsilon' + (3-2n) a_{n-1/2}' \epsilon\;,\\
\delta_n^B \psi_{3/2} &= (n-1/2)\psi_{n-1/2} \eta' + (n-1) \psi_{n-1}' \eta +BF_{n,3/2}\;,\\
\delta_{n-1/2}^B \psi_{3/2} &= \psi_n \eta\;,\\
\delta_n^F \psi_{3/2} &= (2n-2) a_{n-1/2} \epsilon' +(2n-3) a_{n-1/2}' \epsilon\;,\\
\delta_{n-1/2}^F \psi_{3/2} &= 2 a_n \epsilon + FF_{n-1/2, 3/2}\;,
\end{align}
\end{subequations}
where $n \in {\bz}$ and $BF_{i,j}, FB_{i,j}$ and $FF_{i,j}$ represent the non-linear terms, whose explicit forms are given in Appendix \ref{nonlinear}.
The results \eqref{BBhigher} and \eqref{FBhigher} correspond to the conditions that $a$'s and $\psi$'s are primary fields at least at linear order.

Finally, we present the variations including $s = 5/2$ and $s=3$. The bosonic variations of $a$'s are
\begin{eqnarray}
\delta^B_{5/2} a_{5/2} &=& \frac{1-4\lambda}{3}\left[ 2 a_{5/2} \eta' + a_{5/2}' \eta \right] - N^B_{5/2} \left[ 2 a_2 \eta' + a_2' \eta \right] + \frac{k_{CS} N^B_{5/2}}{4 \pi} \eta'''\;,  \label{var5252} \\
\delta^B_{5/2} a_3 &=& 3 a_{7/2} \eta' + a_{7/2}' \eta + \frac{1 - 4 \lambda}{15} \left[ 3 a_3 \eta' + a_3' \eta \right] + BB_{5/2,3}\;,\label{523}\\
\delta^B_3 a_3 &=& 4 a_4 \eta' + 2 a_4'\eta - \frac{N_3^B}{12} \left[ 2 a_2''' \eta + 9a_2'' + 15 a_2' \eta'' + 10 a_2 \eta'''\right]\nonumber\\
 & & + \frac{1 - 4 \lambda}{60} \left[ 2 a_{5/2}''' \eta + 9a_{5/2}'' + 15 a_{5/2}' \eta'' + 10 a_{5/2} \eta'''\right] + \frac{k_{CS} N_3^B}{48} \eta''''' + BB_{3,3}\;,\nonumber\\
\end{eqnarray}
where $BB_{i,j}$ are the non-linear terms and the explicit forms are given in Appendix \ref{nonlinear}. The bosonic variations of $\psi$'s are
\begin{eqnarray}
\delta_{5/2}^B \psi_{5/2} &=& \psi_4 \eta + \frac{1-4\lambda}{15} \left[ 5 \psi_{5/2} \eta' + 2 \psi_{5/2}' \eta \right] - \frac{N_3^B}{12} \left[ 6 \psi_2 \eta'' + 4 \psi_2' \eta' + \psi_2'' \eta \right] + BF_{5/2,5/2}\;,\nonumber\\
&&\\
\delta_{5/2}^B \psi_3 &=& \eta \psi_{7/2} + \frac{1- 4\lambda}{15} \left[ 2 \eta \psi_3' + 5 \eta' \psi_3 \right] - \frac{N^B_3}{12} \left[ \eta \psi_{3/2}'' + 4 \eta' \psi_{3/2}' + 6 \eta'' \psi_{3/2} \right] + BF_{5/2,3}\;,\nonumber\\
&&\\
\delta_3^B \psi_3 &=& \frac{7}{2}\eta' \psi_4 + 2 \eta \psi_4' - \frac{1-4\lambda}{30} \left[ 2 \eta \psi_{5/2}'' + 6 \eta' \psi_{5/2}' + 5 \eta'' \psi_{5/2} \right]  \nonumber\\
& &-\frac{N_3^B}{24} \left[ 4 \eta \psi_2''' + 15 \eta' \psi_2'' + 20 \eta'' \psi_2' + 10 \eta''' \psi_2 \right] + BF_{3,3}\;.
\end{eqnarray}
The fermionic variations of $\psi$'s are
\begin{eqnarray}
\nn \delta^F_{5/2} \psi_{5/2} &=& 2 a_4 \epsilon + \frac{1 - 4 \lambda}{15} \left[ 3 a_{5/2}'' \epsilon + 10 a_{5/2}' \epsilon' + 10 a_{5/2} \epsilon'' \right]\nonumber\\
& &-\frac{N_3^B}{6} \left[ 3 a_2'' \epsilon + 10  a_2' \epsilon' + 10 a_2 \epsilon'' \right] + \frac{k_{CS} N_3^B}{12 \pi} \epsilon'''' + FF_{5/2,5/2}\;,\\
\nn\delta^F_{5/2} \psi_3 &=& -3 a_{7/2}' \epsilon - 6 a_{7/2} \epsilon' + \frac{2(1-4\lambda)}{15} \left[ a_3' \epsilon + 2 a_3 \epsilon' \right] \nonumber\\
& &+ \frac{N^B_3}{12} \left[ a_{3/2}''' \epsilon + 4 a_{3/2}'' \epsilon' + 6 a_{3/2}' \epsilon'' + 4 a_{3/2} \epsilon''' \right] + FF_{5/2,3}\;,\\
\nn\delta_3^F \psi_3 &=& -2 a_4 \epsilon - \frac{1-4\lambda}{15} \left[3a_{5/2}'' \epsilon + 10 a_{5/2}' \epsilon' + 10 a_{5/2} \epsilon'' \right]\\
& &+ \frac{N_3^B}{6} \left[3a_2'' \epsilon + 10 a_2' \epsilon' + 10 a_2 \epsilon'' \right] - \frac{k_{CS} N_3^B \epsilon''''}{12 \pi} + FF_{3,3}\;.
\end{eqnarray}

Note that the variation \eqref{var5252} does not have any non-linear terms, it can be converted into a commutator of the algebra after the shift \eqref{modeshift}
\begin{equation}\label{5252}
[(a_{\frac{5}{2}})_m,~(a_{\frac{5}{2}})_n]=\frac{1-4\l}{3}(m-n)(a_{\frac{5}{2}})_{m+n}-N^{B}_{5/2} (m-n)(a_{2})_{m+n}-\frac{k_{CS}N^{B}_{5/2} }{2}(m^{3}-m)\del_{m+n,0}
\end{equation}
This plays an important role in section \ref{check}, where we compare it with the commutator in the dual CFT.

The super-$W_{\infty}[\l]$ we have just obtained is non-linear due to the non-vanishing curvature of AdS$_3$. As in the bosonic case \cite{Bowcock:1991zk, gh}, these non-linear terms drop once the curvature is taken to zero, and the super-$W_{\infty}[\l]$ algebra further reduces to the $shs[\l]$ algebra if one takes its wedge subalgebra. To see this, one first converts the shift in mode \eqref{modeshift} back to a shift in the energy-momentum tensor according to \eqref{adsmode}:
\begin{equation}
a_{2} \to a_{2}-\frac{k_{CS}}{8 \pi}\,.
\end{equation}
Applying this to the variations of the super-$W_{\infty}[\l]$ algebra will generate some linear terms from the nonlinear terms. The remaining nonlinear terms are negligible in the vanishing curvature limit. The mode expansion according to \eqref{adsmode} then takes the linear terms back to the $shs[\l]$ algebra\fn{Here we do not consider the central terms in the super-$W_{\infty}$ algebra. } (e.g. \eqref{cmfm}).

\section{Identification with the \cpn~chiral algebras}\label{check}

In this section, we examine the duality between $\cn = 2$ higher-spin supergravity and $\cn = 2$ \cpn~model at large $n$ \cite{chr} from the perspective of the symmetry. The authors of \cite{chr} proposed in their work that $\cn = 2$ higher-spin supergravity based on $shs[\lambda] \times shs[\lambda]$ algebra is equivalent to the large-$n$ limit of the $\cn = 2$ Kazama-Suzuki type coset model \eqref{ksmd} (known as \cpn~model \cite{ks}) with
\begin{equation}
n,k \to \infty\;,\qquad \lim\limits_{n,k\to \infty}\frac{n}{2(n+k)} = \lambda\;, \qquad c_{\text{AdS}} = c_{\text{CFT}}\;.
\label{parameterID}
\end{equation}
One can check that the relation between the central charge and the Chern-Simons level is consistent with the asymptotic super-Virasoro algebra (See \eqref{Virasoro}, for example.). The central charge of the coset model $c_{\text{CFT}}$ is known to be $3nk/(n+k+1)$, so after taking the 't Hooft limit, one obtains the identification
\begin{equation}
k_{CS} = \frac{n ( 1 - 2 \lambda)}{2}\;.
\label{parameterID2}
\end{equation}

The goal of this section is to check this duality by understanding the underlying symmetries. The global symmetry of the higher-spin supergravity forms the super-$W_{\infty}[\lambda]$ algebra we obtained in the previous section, and the procedure we followed to get the super-$W_{\infty}[\lambda]$ algebra coincides with the classical Drinfeld-Sokolov (CDS) reduction of the $shs[\lambda]$ algebra. On the other side of the duality, we consider  the large-$n$ limit of the  chiral algebra $\cs\cw_{n}$ of the $\cn=2$ \cpn~model which comes from the quantum Drinfeld-Sokolov (QDS) reduction of the Lie superalgebra $A(n,n-1)$ \cite{Kac:1977em, Dobrev:1985qz, itoqr}. We propose that the $\cs\cw_{n}$ algebra in the large-$n$ limit coincides with the super-$W_{\infty}[\lambda]$ with the parameter identifications \eqref{parameterID}. In the following two subsections, we carry out two non-trivial checks to support the above proposal. We first check the matching of the two algebras and then the matching of the representations of the two algebras.

\subsection{Large $n$ limits of the \cpn~chiral algebra
}
In this section, we match the two algebras by explicitly showing that the variation of the higher-spin fields $a_{\frac{5}{2}}$ under the asymptotic symmetry transformation agrees with the OPE of the corresponding operators in the 't Hooft limit of the \cpn~model. This non-trivial check partially supports the claim that the two algebras are identical.

Before the actual check, we briefly review the chiral algebra structure of the \cpn~minimal model. The chiral algebra $\cs\cw_{n}$ can be derived from $A(n,n-1)$ by the QDS reduction \cite{itoqr}. Concretely, the higher-spin currents can be obained from the super-Lax operator
\begin{equation}\label{lax}
L(Z)\equiv~:(a D-\Theta_{2n+1}(Z))(a D-\Theta_{2n}(Z))\cdots (a D-\Theta_{1}(Z)):\,,
\end{equation}
where $Z=(z,\q)$ is the $\cn=1$ superspace coordinate with $z$ being bosonic and $\q$ being fermionic, $D=\frac{\pa}{\pa \q}+\q\frac{\pa}{\pa z}$ is the super-covariant derivative, $:~:$ denotes the normal ordering, $a$ is a bosonic parameter from the QDS reduction and $\Theta_{i}(Z)=(-1)^{i-1}(\L_{i}-\L_{i-1},~ D\Phi(Z))$. Here, $\L_{i}$ is a fundamental weight of $A(n,n-1)$ with $\L_{0}=0=\L_{2n+1}$, $\Phi$ is a free chiral superfield, taking values in the root space of $A(n,n-1)$ and $(\cdot\,,\cdot)$ represents the inner product on the root space. One can expand $L(Z)$ in terms of $aD$ by moving $aD$ to the very right of the expression, then the coefficients of different powers of $aD$ are the generators of the super-$W_{n}$ algebra \cite{itocp}:
\begin{equation}\label{hscurrent}
L(Z)=(aD)^{2n+1}+
\sum_{i=2}^{2n+1}W_{{i\over 2}}(Z)(a D)^{2n+1-i}\;.
\end{equation}
The superfields $W_{k}$ decompose into components fields. We present here how the first few $W_{k}$'s are decomposed:
\begin{eqnarray}
\nn W_{1}(Z)&=& W_{1}^{-}(z)+i \theta [G_{2}^{+}(z)+G_{2}^{-}(z)]\;, \\
\nn W_{{3\over 2}}(Z)&=& a [i G_{2}^{-}(z)+\theta W_{2}^{+}(z)] \;, \\
\nn W_{2}(Z)&=& W_{2}^{-}(z)+i\theta [G_{3}^{+}(z)+G_{3}^{-}(z)]\;, \\
\nn W_{{5\over 2}}(Z)&=& a [i G_{3}^{-}(z)+\theta W_{3}^{+}(z)]\;,
\end{eqnarray}
where $W_{i}^{\pm}$ are bosonic generators with conformal weight $i$ and $G_{i}^{\pm}$ are fermionic generators with conformal weight $i-\frac{1}{2}$.  We identify $W_{1}^{-}$ and $W_{2}^{+}$ with the familiar $U(1)$ charge and the energy momentum tensor $J$ and $T$ respectively. In addition to the matching of the central charge, we can match the higher spin fields on the both sides of the duality now. The dictionary between the higher-spin fields in the asymptotic algebra and the primaries in the \cpn~model is:
\begin{equation}
a_{s} \leftrightarrow W_{s}^{+}\;, \quad a_{t+1/2} \leftrightarrow W_{t}^{-}\;, \quad \psi_{s} \leftrightarrow G_{s}^{+}\;, \quad \psi_{t+1/2} \leftrightarrow G_{t+1}^{-}\nonumber
\end{equation}
\begin{equation}\label{dictionary}
s,t\in \bz, \quad s\geq 2\;,\quad t\geq 1\;.
\end{equation}
where the spin of the fields in the AdS side is matched with the conformal weight of the operators in the CFT side.

The OPEs between these operators can be computed from the free field realization of $\cs\cw_{n}$ algebra \cite{itocp,itoffr}, where the results are explicitly known  for $n=3$. Computing the OPEs in the 't Hooft limit requires the knowledge of the OPEs at general $n$, which is complicated in general. Our strategy is to compute the OPEs at several small $n$ first and then extrapolate to results at general $n$. However, this extrapolation is possible in principle but difficult in practice, because there are non-linear terms in the OPE between higher spin operators. These non-linear terms make the extrapolation to general $n$ difficult. Nevertheless, in the supersymmetric setting, there are examples such as the $W_{2}^{-}W_{2}^{-}$ OPE that is linear. This makes the general $n$ extrapolation straightforward. For this reason, we restrict our attention to the $W_{2}^{-}W_{2}^{-}$ OPE.

Since our goal is to compare the algebra in the AdS side and that in the CFT side, we need to redefine the operators in the CFT side in such a way that the higher spin operators in the CFT are in the same bases as the ones in the AdS side. Up to the $W_2^-$ operator, this can be done as follows:
\begin{enumerate}
\item Compute the relevant OPEs for $n=2,3,4,5$ and extrapolating the results to the general $n$ expressions.
\item Redefine the super-Virasoro operators, $T$ and $G_{2}^{\pm}$ to $\wt T$ and $\wt{G}_{2}^\pm$ so that the OPEs between the redefined operators match the variations on the AdS side \eqref{supervirasoro}.
\item Redefine the $W_{2}^{-}$ operator so that its OPEs with the operators $J$ and the $\wt T$ match with the corresponding variation on the AdS side \eqref{JBhigher} and \eqref{BBhigher}.
\end{enumerate}
The redefinitions are explicitly given by
\begin{subequations}\label{modall}
\begin{align}
T &\to \wt{T}=T-\frac{1}{2} \pa J-\frac{:JJ:}{2 n \left(1+a^2+n a^2\right)}\\
W_2^- &\to \widetilde{W}_2^-=W_2^- +\frac{1}{2} \left(a^2-n a^2\right) \pa J+\frac{(1-n) :JJ:}{2 n}-\frac{\left(1+a^2+n^3 a^4-n \left(1+a^2+a^4\right)\right) \tilde T}{-1+3 n^2 a^2+3 n \left(1+a^2\right)}
\end{align}
\end{subequations}
Now we carry out the OPE between the modified operators $\widetilde {W}_{2}^{-}$ on the CFT side
\begin{eqnarray}
\nonumber  \wt{W}_{2}^{-}(z)\wt{W}_{2}^{-}(w) &=& -\frac{(-1+n) n \left(-1+a^2 n\right) \left(1+a^2+a^2 n\right) \left(2+a^2+a^2 n\right) \left(1+2 a^2+a^2 n\right)}{2 \left(-1+3 n+3 a^2 n+3 a^2 n^2\right)(z-w)^4} \\
\nonumber   &-&
   \frac{2 (1+n) \left(1+a^2 n\right) \left(1+a^2+2 a^2 n\right)~\wt{W}_{2}^{-}}{\left(-1+3 \left(1+a^2\right) n+3 a^2 n^2\right)(z-w)^2}\\
\nonumber   &-&\frac{2 (-1+n) n \left(-1+a^2 n\right) \left(1+a^2+a^2 n\right) \left(2+a^2+a^2 n\right) \left(1+2 a^2+a^2 n\right) \wt{T}}{\left(-1+3 \left(1+a^2\right) n+3 a^2 n^2\right)^2(z-w)^2}\\
   &-&\frac{(1+n) \left(1+a^2 n\right) \left(1+a^2+2 a^2 n\right) \pa \wt{W}_{2}^{-}}{\left(-1+3 \left(1+a^2\right) n+3 a^2 n^2\right)(z-w)}\\
\nonumber   &-&\frac{n (-1+n) \left(-1+a^2 n\right) \left(1+a^2+a^2 n\right) \left(2+a^2+a^2 n\right) \left(1+2 a^2+a^2 n\right) \pa \wt{T}}{\left(-1+3 \left(1+a^2\right) n+3 a^2 n^2\right)^2(z-w)}\;.
\end{eqnarray}
According to the duality proposed in \cite{chr}, we take the 't Hooft limit \eqref{parameterID}:
\begin{equation}\label{tempthdef}
 n,k \to \infty\;, \qquad \lim\limits_{n,k\to \infty}n a^2= - \lim\limits_{n,k\to \infty} \frac{n}{n+k+1} = - 2 \lambda\;.
\end{equation}
then the above OPE can be rewritten as
\begin{eqnarray}
\nonumber \wt{W}_{2}^{-}(z)\wt{W}_{2}^{-}(w)\bigg|_{\text{'t Hooft limit}}\!\!\!\!&=& -\frac{ (2 \lambda + 1 ) (2 \lambda - 1) ( 1 - \lambda ) n}{3(z-w)^4}\\
\nonumber&~&-\frac{2(3(1- 4 \lambda) \wt{W}_{2}^{-}+2(2 \lambda + 1) ( \lambda - 1 ) \wt{T})}{9(z-w)^2}\\
&~&-\frac{3 (1-4 \lambda) \pa \wt{W}_{2}^{-}+ 2 (2\lambda + 1) ( \lambda - 1 ) \pa \wt{T}}{9(z-w)}\;,\label{CFTope}
\end{eqnarray}
where we keep only the leading term at large $n$.

The OPEs in the CFT are functions of complex variables $z$ and $w$, while the variations of higher-spin fields under the \as~are functions of variable $\q$, so we cannot compare them directly. Therefore, we first convert the results on the both sides to commutators between modes. The converting in the AdS side is given in \eqref{adsmode} and the discussion there. The mode expansion on the CFT side is defined by:
\begin{equation}
\wt{W}=\sum\limits_{n\in\bz} \wt{W}_{n} z^{-n-h_{\wt{W}}}
\end{equation}
where $h_{\wt{W}}$ is the conformal weight of $\wt{W}$. Plugging this into the OPE \eqref{CFTope} and redefining $\wt{W}_2^- \to - \wt{W}_2^-$ yield
 \begin{eqnarray}
\nn[\wt{W}^{-}_{2,m},~\wt{W}^{-}_{2,n}]&=&\frac{(1+2\l) (1-2\l) (1-\l)  n}{18} (m^{3}-m) \del_{m+n,0}\\
&&+\frac{2(1+2\l) (1-\l) }{9} (m-n) \wt{T}_{m+n}+\frac{(1-4\l)}{3}  (m-n) \wt{W}^{-}_{2,m+n} \label{cftmode}
\end{eqnarray}
From the dictionary \eqref{dictionary}, we expect that this commutation relation should match with the $[(a_{\frac{5}{2}})_{m},~(a_{\frac{5}{2}})_{n}]$ commutator \eqref{5252} in the asymptotic symmetry algebra.
Using the relation between the central charge and the Chern-Simons level \eqref{parameterID2} and $N^{B}_{5/2} = (2/9)(-1+\lambda) (1+2\lambda)$, one finds that this commutator exactly agrees with the one in the CFT side \eqref{cftmode}, including the numerical coefficients!

This computation is possible only in the supersymmetric case, since in the bosonic case, the OPE between any pair of higher-spin generators contains non-linear terms and the large-$n$ extrapolation is difficult as discussed above. However, in the supersymmetric case, there exists an OPE \eqref{CFTope} that is linear and the extrapolation is straightforward to carry out. Note that the generator $\wt{W}_{2}^{-}$ is introduced by the $\cn=2$ supersymmetry so the possibility of this check is brought to us by introducing supersymmetry.

\subsection{Degenerate representations}

In this section, we compare the degenerate representations, whose Verma modules are truncated by null vectors, on both AdS and CFT sides. We show that for any degenerate representation whose highest weight state is a chiral primary state can be a degenerate representation of the asymptotic symmetry algebra of the higher-spin supergravity theory and vice versa.

Finding the degenerate representations on the CFT side is straightforward. The chiral algebra of the $\cn=2$ \cpn~minimal model is shown to be the $\cs\cw_{n}$ algebra that can be derived from the Lie superalgebra $A(n,n-1)$ by the QDS reduction \cite{itocp}. We can find the degenerate representations by explicitly constructing the null vectors in the modules and the resulting expressions for the degenerate representations are known (see e.g. \cite{bs,itocp} and reference therein).  We then take the 't Hooft limit by simply applying the limit \eqref{parameterID} to the representations.

The degenerate representations of the asymptotic algebra are not explicitly known in the literature. We thus take a step back and find them indirectly.  First, note that the way we get the asymptotic super-$W_{\infty}[\l]$ algebra on the AdS side is the same as the classical Drinfeld-Sokolov reduction of $shs[\lambda]$. Secondly, we utilize the fact that $shs[\lambda]$ is realized by analytically continuing $A(n,n-1)$ to $n = -2\lambda$ \cite{fl,gghr}. Thirdly, we know that the Quantum Drinfeld-Sokolov reduction of $A(n,n-1)$ gives $\cs\cw_{n}$ algebra and we know how to find its degenerate representations. Finally, one can take the classical limit that reduces the Quantum Drinfeld-Sokolov reduction to the Classical Drinfeld-Sokolov reduction. This limit corresponds to taking the level of QDS reduction, $k_{DS}$, to infinity. Thus, we can start with any degenerate representation of the algebra $\cs\cw_{n}$ and apply the combination of these operations: $n \to -2\lambda, k_{DS}\to \infty$,  then the resulting representation is  a degenerate representation of the super-$W_{\infty}[\l]$ algebra.

With this reasoning in mind, we can compare the degenerate representations on both sides by starting with any degenerate representation of $\cs\cw_{n}$, taking the two limits, (i) Super-higher-spin limit: $n = -2\lambda, k_{DS}\to \infty$ and (ii) the 't Hooft limit: $n,k\to \infty, \lim\limits_{n\to \infty}\frac{n}{n+k+1}=2\l$. Then we compare the spectra of conformal weights and the $U(1)$ charges of the two resulting representations.  The relation is clear in the following diagram\\
[12mm]
\begin{picture}(0,0)(-70,25)
\put(-23,52){$A(n,n-1)$}
\put(105,52){$shs[\l]$}
\put(0,17){$\cs\cw_n$}
\put(100,17){super-$W_\infty[\l]$}
\put(48,58){\scriptsize $n\to-2\l$}
\put(46,15){\scriptsize $n\to-2\l$}
\put(44,8){\scriptsize
$k_{DS}\to \infty$}
\put(38,23){\scriptsize classical limit}
\put(40,54){\vector(1,0){58}}
\put(35,21){\vector(1,0){58}}
\put(13,37){\scriptsize QDS}
\put(115,37){\scriptsize CDS}
\put(10,46){\vector(0,-1){16}}
\put(110,46){\vector(0,-1){16}}
\put(186,17){$\cs\cw_n$}
\put(272,17){super-$W_\infty[\l]$}
\put(218,20){\vector(1,0){50}}
\put(218,25){\scriptsize 't Hooft limit}
\put(238,16){\small $?$}
\put(205,50){\small AdS$_3$/CFT$_2$ Proposal}
\multiput(176,0)(0,4){18}{\line(0,1){2}}
\end{picture}\\
[1mm]

Let us now move on to computing the degenerate representations. We start from any degenerate representation of $\cs\cw_{n}$. In the bosonic sector, the highest weight state of the module is characterized by a weight of the form:
\begin{equation}\label{degrep}
    \Lambda=\a_+ \Lambda_++\a_- \Lambda_-\;,
\end{equation}
where $\a_-=-\sqrt{k_{DS}+1}$ and $\a_-\a_+=-1$. $\Lambda_+$ and $\Lambda_-$ are linear combinations of fundamental weights with non-negative integer coefficients.
For a given $\Lambda$, the conformal  dimension is represented as \cite{itoqr, bs}
\begin{equation}\label{confweight}
h(\Lambda) = \frac{1}{2} (\Lambda, \Lambda + 2 \alpha_- \r)\;,
\end{equation}
where $\r$ is the dual Weyl vector \cite{pc}. The $U(1)_R$ charge is given by
\begin{equation}
Q(\Lambda) = - \a_- (\Lambda, \nu)\;,
\end{equation}
where $\nu$ is the generator of the center of the $A(n,n-1)$ algebra, whose expression is given explicitly in Appendix C.

We first consider the spectrum in the CFT side. To find out the relation between $k_{DS}$ and the level in the coset model $k$, we match the central charge of $\cs\cw_{n}$ from the QDS reduction with that of the \cpn~coset model \cite{itoqr,itocp,bs}. It yields
\begin{eqnarray}
 \nonumber  c_{mm} &=& \frac{3nk}{n+k+1} \\
 \nonumber  c_{DS} &=& 3n(1-(n+1)\a_-^2) \\
 c_{mm}= c_{DS} \Rightarrow k_{DS}+1&=& \frac{1}{n+k+1} \;,\label{kdsandk}
\end{eqnarray}
where the subscript $mm$ stands for the minimal model. Therefore, $k_{DS} + 1 \to 0$ ($\a_-\to 0^-$ and $\a_+\to\infty$) in the 't Hooft limit. To extract the representations with finite conformal dimensions, one needs to set $\Lambda_+=0$. Then we get
\begin{equation}
h_{mm}(\Lambda) = \frac{1}{2} \a_-^2 ( \Lambda_-, \Lambda_- + 2 \r)\;.
\end{equation}
This can be evaluated and re-expressed in terms of the quantities associated with Young superdiagram for $A(n,n-1)$. Using the definition of the 't Hooft limit \eqref{parameterID}, \eqref{ynglr}, one obtains
\begin{equation}\label{CFTweight}
h(\Lambda) = \lambda( B - \bar{B} )\;,
\end{equation}
\begin{equation}\label{CFTcharge}
q( \Lambda ) = 2 \lambda ( B - \bar{B} )\;,
\end{equation}where $B$ is the number of boxes in the covariant part of the Young supertableaux, $\bar{B}$ is that in its contravariant part and $B-\bar{B}$ can take any non-negative integer values.\fn{One can show that the weight $\L_{\pm}$ being integral dominant guarantees that $B-\bar{B}\geq 0$. See \eqref{BminusbarB}.}

The fermionic sector represents the affine Lie algebra $SO(2n)$ at level one. The degenerate representation is characterized by a weight $\tilde\L$ and the contributions to the conformal weight and the $U(1)_R$ charge are $\frac{1}{2} \tilde \L^2$ and $\sum_i \tilde{\Lambda}_i$, respectively \cite{ks}. Therefore, the conformal weight of a degenerate representation ends up with
\begin{equation}
h_{mm}(\Lambda) = \lambda( B - \bar{B} ) + \frac{1}{2} \tilde \L^2\;,
\end{equation}
and the $U(1)_R$ charge with
\begin{equation}
q_{mm}( \Lambda, \tilde{\Lambda} ) = 2 \lambda ( B - \bar{B} ) + {\displaystyle \sum_i} \tilde{\Lambda}_i\;.
\end{equation}
To obtain chiral primaries, one needs to set $\tilde{\Lambda} = 0$ \cite{Gepner:1988wi}. Then, the condition for chiral primary, $h=\frac{1}{2}q$, is automatically satisfied. Hence, the spectrum of chiral primaries is given by
\begin{equation}
h_{mm}( \Lambda, \tilde{\Lambda} ) = \lambda( B- \bar{B} )\;.\label{CPmm}
\end{equation}

Now we want to compute the conformal weights and $U(1)_R$ charge of the degenerate representation of the symmetry algebra in the AdS side. To get that, one takes $n\to -2\l,~k_{DS} \to \infty$. In this limit, $\a_-$ diverges, and one needs to set $\Lambda_- = 0$ to get representations with finite conformal weights. The conformal weight can be evaluated using \eqref{ynglr} as
\begin{eqnarray}
h_{HS}(\Lambda, \tilde{\Lambda}) &=& - (\Lambda_+, \rho)+ \frac{1}{2} \tilde{\Lambda}^2\nonumber\\
&=& \l B +\frac{1}{2}(2\l-1) \bar{B}-\frac{1}{2}\cb-\frac{1}{2}\sum\limits_{i}\bar{r}_i^2 +\frac{1}{2}\sum\limits_{i}c_i^2+ \frac{1}{2} \tilde{\Lambda}^2 \;.
\end{eqnarray}
The $U(1)_R$ charge is evaluated using \eqref{yngln} as
\begin{eqnarray}
q_{HS}(\Lambda, \tilde{\Lambda}) &=&  -\a_- (\a_+\L_+, \n) + \sum\limits_i \tilde{\Lambda}_i\\
&=& 2\l B+(1-2\l)\bar B+ \sum\limits_i \tilde{\Lambda}_i\;. \label{qds}
\end{eqnarray}
As in the 't Hooft limit, one needs to impose $\tilde{\Lambda}=0$ to obtain chiral primaries. In addition, the chiral primaries should satisfy the condition $h = \frac{1}{2}q$. This condition should hold independently for $\lambda$-dependent and independent parts, and the consideration of the $\lambda$-independent part yields $\bar{B} = 0$, which further results in $\sum\limits_{i}\bar{r}_i^2 = 0$. Then, if we subtract $\frac{1}{2}q$ from $h$, we obtain $\frac{1}{2} ( - B + \sum\limits_{i}c_i^2)$, which is non-negative and vanishes only when $c_i=1$.\fn{This means the corresponding Young superdiagram has only one row.}  Therefore, the conformal weights for the chiral primaries are given by
\begin{equation}
h_{HS}(\Lambda, \tilde{\Lambda}) = \lambda B\;.\label{CPHS}
\end{equation}
Note that $B$ can take any non-negative integer value. By comparing \eqref{CPmm} and \eqref{CPHS}, we find that the spectra of chiral primaries exactly agree on both limits by identifying $B-\bar{B}$ in the 't Hooft limit and $B$ in the higher-spin limit. This agreement provides further evidence that the 't Hooft limit of the $\cs \cw_n$ algebra is equivalent to the super-$W_\infty[\lambda]$ algebra we obtained in the previous section. The agreement of the spectrum of general complete degenerate representations is not quite obvious, we hope to get back to this problem in later study.

This indicates that the representation in the 't Hooft limit of the \cpn~model can be a representation of the asymptotic $\cs \cw_{\infty}$ algebra on the AdS side and vice versa. Thus this matching of chiral primary representations on the two sides provides another piece of evidence for the validity of the duality.

\section{Conclusion}

In this paper, we have analyzed the asymptotic symmetry of the supergravity theory supersymmetrically coupled to an infinite tower of higher-spin fields. The matching of this asymptotic symmetry algebra with the chiral algebra of the \cpn~CFT model in the 't Hooft limit provides another non-trivial check of the recently proposed supersymmetric duality \cite{chr}. We have also found that the chiral primaries on both sides of duality have the same spectrum.

For future directions, it would be interesting to extend the matching of the symmetry algebras to higher order. Due to the technical difficulties, we found it hard to obtain the commutators for higher-spin generators in the coset CFT for general $n$. It is, however, possible in principle, and should provide firmer evidence for the duality. Another direction is to compute the partition function and correlation functions on both sides and see the agreement so as to provide other strong evidence for the duality. The one-loop partition function was discussed in the recent paper \cite{Candu:2012jq}.

\section*{Acknowledgements}
We are grateful to Henriette Elvang, Michael Gutperle, Kentaro Hori, Katsushi Ito and Finn Larsen for useful and enlightening discussions. We would like to thank especially Tom Hartman for illuminating discussion and helpful comments on the draft.
We appreciate the comments from Changhyun Ahn and Constantin Candu and their suggestions for the revised version of this paper. CP is supported by NSF Grant PHY-0953232. KH and CP are supported in part by the DOE Grant DE-FG02-95ER 40899.

\appendix

\section{Super-higher-spin algebra $shs[\l]$}
\label{shsapp}

We briefly summarize the basic facts about the $shs[\l]$ algebra. The super-higher-spin algebra is generated by bosonic generators $L_{m}^{(s)\pm}$ as well as fermionic generators $G_{r}^{(s)\pm}$. It can be obtained as the wedge subalgebra of the super-$W_{\infty}[\l]$ algebra constructed in \cite{bdv2}.\fn{Note that this super-$W_{\infty}(\l)$ algebra in \cite{bdv2} is not the asymptotic algebra we found in the main text, although accidentally they have the same name.} The wedge is taken to be:
\begin{equation}
|m|\leq s-1,\qquad|r|\leq s-\frac{3}{2}\,.
\end{equation}
This wedge condition restricts the generators with given spin $s$ to be in finite-dimensional irreducible representations\fn{${\text {dim}}(L^{(s)})=2s-1, {\text {dim}}(G^{(s)})=2s-2,~s\in \bz$} of the bosonic $sl(2,\br)$ algebra. A realization of these operators as differential operators in $\cn=1$ superspace is given in \cite{bdv2}.

The following definition of the $shs[\l]$ algebra is convenient for our later discussion. Consider the universal enveloping algebra of  $Osp(1,2)$ factered out an ideal $\c$:
\begin{equation}\label{uea}
SB[\l]=U(Osp(1,2))/\c\,,\quad~~~~\c=\<~C_{2}(Osp(1,2))-\l(\l-\frac{1}{2})~\>\,,
\end{equation}
where $C_{2}(Osp(1,2))$ is the quadratic Casimir of $Osp(1,2)$. The super-higher-spin algebra (as a vector space) is identified with a subspace of $SB[\l]$:\begin{equation}\label{shs}
    U(Osp(1,2))/\c=shs[\l]\oplus\bc\,.
\end{equation}
The $\bc$ is generated by the identity element $\bf 1$, which corresponds the $L^{(1)+}_0$ generator in the realization \cite{bdv2}. $U(Osp(1,2))$ is an associative algebra. We denote (associative) multiplication between two elements in $U(Osp(1,2))$ as $\cj\star\cl$. The commutator of $shs[\l]$ is defined by the multiplication in $U(Osp(1,2))$:
\begin{equation}\label{commtt}
  [\cj,\cl]=\cj\star\cl-\cl\star\cj\,.
\end{equation}
In our computation, we further define a bilinear trace of the product of two elements in $SB[\l]$:
\begin{eqnarray}\label{trace}
&&\mbox{Tr}(\ck,\cl)=\frac{\ck\star\cl}{(2\l^{2}-\l)}\bigg|_{\cj=0}, ~~~\forall \cj\neq {\bf 1}
\end{eqnarray}
Note the product $\ck\star\cl$ can be expanded in terms of generators of $SB[\l]$, the right hand side of \eqref{trace} means we keep only terms proportional to $\bf 1$ (or $L^{(1)+}_0$ in the language of \cite{bdv2}) and send all the other generators of $SB[\l]$ to zero. We further divide out a factor of $2\l^{2}-\l$ to make sure that our normalization functions $N^B_s$, $N^F_s$, which are bilinear traces of special pairs of elements in $SB[\l]$
\begin{equation}
N^B_s = \mbox{Tr} (L^{(s)}_{-\fs+1} L^{(s)}_{\fs-1}), \quad~~~~{N}^F_s = \mbox{Tr} (G^{(s)}_{\lceil s\rceil-3/2} G^{(s)}_{\fms+3/2} )\,,
\end{equation}
give the correct value at $s=2$. We present here some examples of the normalization functions that are used in our computation:
\begin{eqnarray}
\nn N^B_{\frac{3}{2}}&=&-2,\hspace{65mm} {N}^F_{\frac{3}{2}}=2\,,\\
\nn N^B_{2}&=&-1,\hspace{65mm} {N}^F_{2}=-2\,,\\
\nn N^B_{\frac{5}{2}}&=&\frac{2}{9} (-1+\lambda ) (1+2 \lambda ),\hspace{36mm} {N}^F_{\frac{5}{2}}=-\frac{2}{3} (-1+\lambda ) (1+2 \lambda )\,,\\
N^B_{3}&=&\frac{2}{3} (-1+\lambda ) (1+2 \lambda ),\hspace{36mm} {N}^F_{3}=\frac{2}{3} (-1+\lambda ) (1+2 \lambda )\,,\\
\nn N^B_{4}&=& \frac{2}{5} (1-\lambda ) (1+\lambda ) (-3+2 \lambda ) (1+2 \lambda )\,,\hspace{8mm} N^F_{4}=\frac{4}{15} (1-\lambda ) (1+\lambda ) (-3+2 \lambda ) (1+2 \lambda )\,.
\end{eqnarray}
The algebraic structure of the $shs[\l]$ algebra is encoded in the following commutation relations:
\begin{eqnarray}
\nn  [L^{(s)}_m,L^{(t)}_n] &=&\sum\limits_{u=1}^{s+t-1} g^{st}_u (m,n,\l) L^{(s+t-u)}_{m+n} \\
\nn  \{G^{(s)}_p,G^{(t)}_q\} &=&\sum\limits_{u=1}^{s+t-1} \tilde{g}^{st}_u (p,q,\l) L^{(s+t-u)}_{p+q}\\
\nn ~ [L^{(s)}_m, G^{(t)}_q] &=&\sum\limits_{u=1}^{s+t-1} h^{st}_u (m,q,\l) G^{(s+t-u)}_{m+q} \\
 ~ [G^{(s)}_p,L^{(t)}_n] &=&\sum\limits_{u=1}^{s+t-1} \tilde{h}^{st}_u (p,n,\l) G^{(s+t-u)}_{p+n} \label{cmfm}
\end{eqnarray}
where $h^{st}_u (m,q,\l)=-\tilde{h}^{ts}_u (q,m,\l)$.
The structure constants $ g^{st}_u (m,n,\l)$, $ \tilde{g}^{st}_u (p,q,\l)$, $h^{st}_u (m,q,\l)$ can be derived from the associate multiplication \eqref{commtt} of $SB[\l]$. Here we first review the commutation relations in \cite{bdv2}, then we show how to get the structure constants in \eqref{cmfm} from the results in \cite{bdv2}. The generators of the $shs[\l]$ algebra can be expressed in $\cn=1$ supersymmetric language as:
\begin{equation}\label{sl}
\cl_{\l}^{(s)}(\O^{(s)})=\sum\limits_{m=1- s}^{s-1} \L_{-m}^{(s)}L_{m}^{(s)}-\sum_{r=\frac{3}{2}-s}^{s-\frac{3}{2}}\Q^{(s)}_{-r}G_{r}^{(s)}
\end{equation}
where $\L^{(s)}_{m}(\Q_{r}^{(s)})$ are Grassmann even (odd) parameters and
\begin{equation}\label{omega}
\O^{(s)}=
\begin{cases} \L^{(s)+}+2\q \Q^{(s)+}\,, ~~\quad\quad\quad s\in \bz \\ \Q^{(s+\frac{1}{2})-} +\q \L^{(s-\frac{1}{2})-}\,,\quad\quad s\in \bz+\frac{1}{2}
\end{cases}
\end{equation}
with expansions $\L^{(s)\pm}=\sum_{n}\L^{(s)\pm}_{n}z^{n+s-1}$ and $\Q^{(s)\pm}=\sum_{r}\L^{(s)\pm}_{r}z^{r+s-\frac{3}{2}}$. Note we can separate each individual mode $L^{(s)}_m ( G^{(s)}_r)$ in \eqref{sl} by setting $\L^{(s')}_{-m'} \to \del_{s,s'}\del_{m,m'}$ $( \Q^{(s')}_{-r'} \to\, \sim \!\del_{s,s'}\del_{r,r'})$. The commutation relation between generators can be computed as follows \cite{bdv2}:
\begin{eqnarray}
&&\big[\cl^{(s)}(\O^{(s)}),\cl^{(t)}(\O^{(t)})\big]=\sum\limits_{u=1}^{s+t-1}\cl^{(s+t-u)}(\x^{(s+t-u)}_{(s)(t)})\\
\nn &&\x^{(s+t-u)}_{(s)(t)}=f^{u}_{st}(\l)\sum_{i=0}^{2u-2}(-1)^{[\frac{i}{2}+2i(s+u)]}\bigg[
\begin{array}{c}
u-1\\
i/2
\end{array}\bigg]([2s-u])_{[u-1-i/2]+|2u|_{2}|2u-2-i|_{2}}\\
&&~~~~~~~~~~~~~\times([2t-u])_{[i/2]+|2u|_{2}|i|_{2}}(D^{i}\O^{(s)})(D^{{2u-2-i}}\O^{t})\,,
\end{eqnarray}
where $u$ takes both integer and half integer values, $i$ takes integer values and:
\begin{eqnarray*}
|n|_{2}&\equiv& n-2[n/2]\,,\\
D&=&\frac{\pa}{\pa \q}-\q\frac{\pa}{\pa z}\,,\\
(a)_{n}&=&a(a+1)(a+2)\ldots(a+n-1),~(a)_{0}=1\,,\\
\bigg[\begin{array}{c}
a\\
b
\end{array}\bigg]&\equiv& \frac{[a]!}{[b]![a-b]!}\,,\\
f^{u}_{st}(\l)&=&F^{u}_{st}(\l)+(-)^{[-u]+4(s+u)(t+u)} F^{u}_{st}(\frac{1}{2}-\l)\,,\\
F^{u}_{st}(\l)&=& (-)^{[s+t-u-1]}\frac{(2s+2t-2u-2)!}{(2s+2t-[u]-3)!}\sum\limits_{i=0}^{2s-2}\sum\limits_{j=0}^{2t-2} \del(i+j-2s-2t+2u+2)\\
&&\times A^{i}(s,\frac{1}{2}-\l) A^{j}(t,\l) (-)^{2s+2i(s+t-u)}\,,\\
A^{i}(s,\l)&=& (-)^{[s]+1+2s(i+1)}\bigg[\begin{array}{c}
s-1\\
i/2
\end{array}\bigg]\frac{([(i+1)/2]+2\l)_{[s-1/2]-[(i+1)/2]}}{([s+i/2])_{2s-1-[s+i/2]}}\,.
\end{eqnarray*}
This form of the commutation relation was derived in \cite{bdv2}. However, it is not convenient for us to use, so we convert the above results to the more familiar form \eqref{cmfm}: the structure constants $ g^{st}_u (m,n,\l)$, $ \tilde{g}^{st}_u (p,q,\l)$, $h^{st}_u (m,q,\l)$ are simply the function $\x^{(s+t-u)}_{(s)(t)}$ with only $\L^{(s)}_{-m}\L^{(t)}_{-n}$, $ \Q^{(s)}_{-p}\Q^{(t)}_{-q} $ and $ \L^{(s)}_{-m}\Q^{(t)}_{-q}$ turned on respectively as discussed below \eqref{omega}.\fn{We need an extra minus sign for $\tilde{g}^{st}_u (p,q,\l)$ due to the ordering of Grassmann variables.}

We give some examples from our computation at small $s,t,u,m,n$:
\begin{eqnarray*}
  g^{2,\frac{3}{2}}_2 (-1,2,\l)\!\!\!\!&=&\!\!\!\!-2, ~~~g^{\frac{5}{2},\frac{3}{2}}_\frac{5}{2} (-1,2,\l) = \frac{4}{3} (-1+4 \lambda ),~~~~~g^{\frac{5}{2},\frac{5}{2}}_3 (-1,\frac{5}{2},\l)=\frac{7}{9} \left(-1-\lambda +2 \lambda ^2\right)\,,\\
  \tilde{g}^{\frac{3}{2},2}_{2}(0,1,\l)\!\!\!\! &=&\!\!\!\! -1 , ~~~ \tilde{g}^{\frac{3}{2},\frac{5}{2}}_{\frac{5}{2}}(0,1,\l)=\frac{1}{6} (1-4 \lambda ),~~~~~~~~~\,\tilde{g}^{\frac{3}{2},\frac{7}{2}}_{3}(0,1,\l)=\frac{1}{4}+\frac{\lambda }{2}-\lambda ^2\,,\\
   h^{2,\frac{3}{2}}_2 (-1,2,\l)\!\!\!\!&=&\!\!\!\!-\frac{5}{2},~~ h^{\frac{5}{2},\frac{5}{2}}_\frac{5}{2} (-1,2,\l)=\frac{7}{15} (-1+4 \lambda ),~~~~h^{\frac{5}{2},\frac{5}{2}}_3 (-1,\frac{5}{2},\l)=-\frac{2}{3} \left(-1-\lambda +2 \lambda ^2\right)\,,\\
     \tilde{h}^{\frac{3}{2},2}_2 (0,1,\l)\!\!\!\!&=&\!\!\!\!-\frac{1}{2},    ~~\tilde{h}^{\frac{3}{2},3}_\frac{5}{2} (0,1,\l)=\frac{1}{12} (-1+4 \lambda ),~~~~~~~\tilde{h}^{\frac{3}{2},\frac{7}{2}}_3 (0,1,\l)=-\frac{3}{40} \left(-1-2 \lambda +4 \lambda ^2\right)\,.
\end{eqnarray*}
The commutation relations of the $shs[\l]$ algebra show that for any integer $N>2$, the generators $L^{(s)}_{m}, G^{(s)}_{r}$ with $s\geq N$ generate a proper subalgebra at the special value $\l=\frac{1-N}{2}$. In addition, the bilinear trace  \eqref{trace} degenerates, 
\begin{equation}
\Tr(L^{(s)}_{m}L^{(t)}_{n})=0\,,\hspace{10mm} \Tr(G^{(s)}_{p}G^{(t)}_{q})=0\,, \qquad \qquad \text{for}  \quad s>N\,.
\end{equation}
This implies that we can consistently set all generators $L^{(s)}_{m}, G^{(s)}_{r}$ with $s> N$ to zero and obtain a finite Lie superalgebra $sl(N,N-1)$.

\section{Nonlinear terms in super-$W_{\infty} [\lambda]$}
\label{nonlinear}

In this appendix, we present the non-linear terms in the super-$W_{\infty}[\lambda]$ algebra obtained in Section 3.2. First, the non-linear terms in the commutators of two bosonic generators are
\begin{eqnarray*}
BB_{5/2, 3} &=& \frac{4 \pi(1-4\lambda)}{15k_{CS}} \psi_{3/2} \psi_{5/2} \eta
- \frac{4 \pi(1-4\lambda)}{15k_{CS}} \psi_2 \psi_3\eta \nonumber\\
& &+ \frac{\pi N_3^B}{6k_{CS}} (\psi_{3/2} \psi_2' - \psi_{3/2}' \psi_2)\eta\,,\\
BB_{3,3} &=& \frac{16 \pi N^B_3}{3k_{CS}} a_2 (a_2 \eta)'
-\frac{16 \pi(1-4\lambda)}{15 k_{CS}}[(a_2 a_{5/2})' \eta + 2 a_2 a_{5/2} \eta'] \nonumber\\
& &+\frac{16 \pi (11+2 \lambda - 4 \lambda^2)}{15 k_{CS} N_3^B} a_{5/2} (a_{5/2} \eta)'\nonumber\\
& &+ \frac{4 \pi^2 (1-4\lambda)}{15 k_{CS}^2} a_{3/2} (\psi_{3/2} \psi_{5/2} + \psi_2 \psi_3) \eta
+ \frac{2 \pi^2 N^B_3}{3 k_{CS}^2} a_{3/2} (\psi_{3/2} \psi_2' - \psi_{3/2}' \psi_2) \eta \nonumber\\
& &+ \frac{11 \pi (1-4 \lambda)}{ 30k_{CS} } ( 2 \psi_{2} \psi_{5/2} \eta' +(\psi_2 \psi_{5/2})' \eta - 2 \psi_{3/2} \psi_{3} \eta' - (\psi_{3/2} \psi_2)' \eta) \nonumber\\
& &+ \frac{7 \pi N^B_3}{ 12k_{CS} } (\psi_{2} \psi_{2}'' - \psi_{3/2} \psi_{3/2}'') \eta
+ \frac{7 \pi N^B_3}{ 6k_{CS} } (\psi_{2} \psi_{2}' - \psi_{3/2} \psi_{3/2}') \eta'\,.
\end{eqnarray*}
Then, the non-linear terms in the commutators of bosonic and fermionic generators are
\begin{eqnarray*}
FB_{2,2} &=&\frac{\pi}{k_{CS}} a_{3/2} \psi_{3/2} \epsilon\;,\\
FB_{5/2,2} &=& \frac{\pi}{k_{CS}} a_{3/2} \psi_3 \epsilon\;,\\
FB_{3,2} &=& \frac{\pi}{k_{CS}} a_{3/2} \psi_{5/2} \epsilon\;,\\
BF_{3,3/2} &=&  \frac{2 \pi}{k_{CS}} (a_{5/2} \psi_2 + a_{3/2} \psi_3) \eta\;,\\
BF_{3,2} &=& \frac{2 \pi}{k_{CS}} (a_{5/2} \psi_{3/2} + a_{3/2} \psi_{5/2})  \eta\;,\\
BF_{5/2,5/2} &=& -\frac{6 \pi (1-4\lambda)}{5k_{CS}} a_{5/2} \psi_2 \eta
+ \frac{2 \pi (1-4\lambda)}{15k_{CS}} a_{3/2} \psi_3 \eta
+ \frac{3 \pi N_{3}^B}{2k_{CS}} a_2 \psi_2 \eta \nonumber\\
&&- \frac{\pi N_3^B}{12 k_{CS}} (4 a_{3/2} \psi_{3/2} \eta' + 2 a_{3/2} \psi'_{3/2} \eta+a_{3/2}' \psi_{3/2} \eta)
- \frac{\pi^2 N_3^B}{12 k_{CS}^2} a_{3/2}^2 \psi_2 \eta\,, \\
BF_{5/2,3} &=& -\frac{6 \pi (1-4\lambda)}{5k_{CS}} a_{5/2} \psi_{3/2} \eta
+ \frac{2 \pi (1-4\lambda)}{15k_{CS}} a_{3/2} \psi_{5/2} \eta - \frac{\pi^2 N_3^B}{12 k_{CS}^2} a_{3/2}^2 \psi_{3/2} \eta\nonumber\\
&&- \frac{\pi N_3^B}{12 k_{CS}} ( a_{3/2}' \psi_2 \eta + 4 a_{3/2} \psi_2 \eta' + 2a_{3/2} \psi_2' \eta)
+ \frac{3 \pi N_3^B}{2k_{CS}} a_2 \psi_{3/2} \eta\,,\\
BF_{3,3} &=&\frac{2 \pi}{k_{CS}} (a_{3/2} \psi_{7/2} + a_{5/2} \psi_{5/2}) \eta
- \frac{2 \pi (1-4\lambda)}{3k_{CS}} a_3 \psi_{3/2} \eta
- \frac{23 \pi (1- 4 \lambda)}{15k_{CS}} a_{5/2} \psi_2 \eta'\nonumber\\
&&- \frac{16 \pi^2 (1-4\lambda)}{15k_{CS}^2} a_{3/2} a_{5/2} \psi_{3/2} \eta
- \frac{\pi^2 (1-4\lambda)}{15k_{CS}^2} a_{3/2}^2 \psi_{5/2} \eta
+ \frac{2 \pi (1-4\lambda)}{3 k_{CS}} a_2 \psi_{5/2} \eta\nonumber\\
&&- \frac{\pi (1-4\lambda)}{5 k_{CS}} a_{3/2} \psi_{3} \eta'
- \frac{16 \pi (1-4\lambda)}{ 15k_{CS} } a_{5/2} \psi_2' \eta
- \frac{2 \pi (1-4\lambda)}{15 k_{CS}} a_{3/2} \psi_3' \eta\nonumber\\
&&- \frac{\pi (1-4\lambda)}{15 k_{CS}} a_{3/2}' \psi_3 \eta
- \frac{11 \pi (1-4\lambda)}{15 k_{CS}} a_{5/2}' \psi_2 \eta
- \frac{\pi^3 N_3^B}{6 k_{CS}^3} a_{3/2}^3 \psi_{3/2} \eta\nonumber\\
&&+ \frac{3\pi^2 N_3^B}{k_{CS}^2} a_{3/2} a_2 \psi_{3/2} \eta
- \frac{5 \pi N_3^B}{6 k_{CS}} a_{3/2} \psi_{3/2} \eta'''
- \frac{5 \pi^2 N_3^B}{8 k_{CS}^2} a_{3/2}^2 \psi_2 \eta'\nonumber\\
&&+ \frac{55 \pi N_3^B}{12 k_{CS}} a_2 \psi_2 \eta'
- \frac{5 \pi N_3^B}{4 k_{CS}} a_{3/2} \psi_{3/2}' \eta'
- \frac{\pi^2 N_3^B}{2 k_{CS}^2} a_{3/2}^2 \psi_2' \eta
+ \frac{3 \pi N_3^B}{k_{CS}} a_2 \psi_2' \eta\nonumber\\
&&- \frac{\pi N_3^B}{2 k_{CS}} a_{3/2} \psi_{3/2}'' \eta
- \frac{5 \pi N_3^B}{8k_{CS}} a_{3/2}' \psi_{3/2} \eta'
- \frac{\pi^2 N_3^B}{2k_{CS}^2} a_{3/2} a_{3/2}' \psi_2 \eta\nonumber\\
&&- \frac{\pi N_3^B}{2k_{CS}} a_{3/2}' \psi_{3/2}' \eta
+ \frac{7 \pi N_3^B}{3k_{CS}} a_2' \psi_2 \eta
- \frac{\pi N_3^B}{6k_{CS}} a_{3/2}'' \psi_{3/2} \eta\,.
\end{eqnarray*}
Finally, the non-linear terms in the commutators of two fermionic generators are
\begin{eqnarray*}
FF_{3,2} &=&  \frac{4 \pi}{k_{CS}}a_{3/2} a_{5/2}\e\;,\\
FF_{5/2,3/2} &=& - \frac{4\pi}{k_{CS}} a_{3/2} a_{5/2} \epsilon\;,\\
&&\frac{\pi  N^B_{3}\left(3 \epsilon {a'}_{{3}/{2}}^2+6 a_{{3}/{2}}^2 \epsilon ''+4 a_{{3}/{2}} \left(3 \epsilon ' a_{{3}/{2}}'+\epsilon  a_{{3}/{2}}''\right)\right)}{12 k_{CS}}+\frac{3 \pi  \epsilon  a_2^2 N^B_{3}}{k_{CS}}+\frac{4 \pi  \epsilon  (1-4 \lambda ) a_{{3}/{2}} a_3 }{15 k_{CS} }\\
&&-\frac{12 \pi  \epsilon  (1-4 \lambda ) a_2 a_{{5}/{2}} }{5 k_{CS}}-\frac{6 \pi  \epsilon a_{{3}/{2}} a_{{7}/{2}}}{k_{CS}}-\frac{\pi  \psi _{{3}/{2}}\psi _{{3}/{2}}'\epsilon  N^B_{3}}{12 k_{CS}}+\frac{3 \pi  \psi _2\psi _2'\epsilon  N^B_3}{4 k_{CS}}\,,\\
FF_{5/2,5/2} &=&
-\frac{6 \pi}{k} a_{3/2} a_{7/2} \epsilon + \frac{2 \pi^2 (1-4\lambda)}{3k_{CS}^2} a_{3/2}^2 a_{5/2} \epsilon
- \frac{12 \pi(1-4\lambda)}{5k_{CS}} a_2 a_{5/2} \epsilon \nonumber\\
&& + \frac{ \pi^3 N_3^B}{12 k_{CS}^3} a_{3/2}^4 \epsilon - \frac{5 \pi^2 N_3^B}{3 k_{CS}^2} a_{3/2}^2 a_2 \epsilon + \frac{3 \pi N_3^B}{k_{CS}} a_2^2 \epsilon +\frac{\pi N_3^B}{4k_{CS}} a_{3/2}' a_{3/2}'\epsilon \nonumber\\
&& + \frac{\pi N_3^B}{3k_{CS}} a_{3/2} a_{3/2}'' \epsilon +\frac{\pi N_3^B}{k_{CS}} a_{3/2} a_{3/2}' \epsilon' + \frac{\pi N_3^B}{2k_{CS}} a_{3/2}^2 \epsilon''+\frac{4 \pi  \epsilon  (1-4 \lambda ) a_{{3}/{2}} a_3 }{15 k_{CS} }\nonumber\\
&&-\frac{\pi N_3^B}{12 k_{CS}} \psi_{3/2} \psi_{3/2}' \epsilon
+\frac{3 \pi N_3^B}{4 k_{CS}} \psi_2 \psi_2' \epsilon+\frac{12 \pi  \epsilon  N^{B}_{4 }a_{{5}/{2}}^2}{k_{CS} (N^B_3)^{2}}\,,\\
FF_{5/2,3} &=&
\frac{2 \pi(1-4\lambda)}{3k_{CS}} (a_{3/2} a_{5/2})' \epsilon + \frac{4 \pi(1-4\lambda)}{3k_{CS}} a_{3/2} a_{5/2} \epsilon'\nonumber\\
&& - \frac{5 \pi N_3^B}{3k_{CS}} (a_{3/2} a_2)' \epsilon
- \frac{10 \pi N_3^B}{3k_{CS}} a_{3/2} a_2 \epsilon' + \frac{\pi^2 N_3^B}{2k_{CS}^2} a_{3/2}^2 a_{3/2}' \epsilon + \frac{\pi^2 N_3^B}{3k_{CS}^2} a_{3/2}^3 \epsilon'\nonumber\\
&&-\frac{17 \pi N_3^B}{12 k_{CS}} \psi_{3/2} \psi'_2 \epsilon + \frac{3 \pi N_3^B}{4k_{CS}} \psi_{3/2}' \psi_2 \epsilon\,, \\
FF_{3,3} &=& -\frac{\pi^3 \epsilon  a_{{3}/{2}}^4 N^B_3}{12 k_{CS}^3}+\frac{5\pi^2 \epsilon  a_{{3}/{2}}^2 a_2 N^B_3}{3 k_{CS}^2}-\frac{2 \pi ^2 \epsilon  (1-4 \lambda ) a_{{3}/{2}}^2 a_{{5}/{2}}}{3 k_{CS}^2}-\frac{4 \pi  \epsilon  N^{B}_{4} a_{{5}/{2}}^2}{k_{CS} (N^B_3)^{2}}\\
&&-\frac{3 \pi  \epsilon  a_2^2 N^B_3}{k_{CS}}-\frac{4\pi  \epsilon  (1-4 \lambda ) a_{{3}/{2}} a_3}{15 k_{CS}}+\frac{12 \pi  \epsilon  (1-4 \lambda ) a_2 a_{{5}/{2}}}{5 k_{CS}}\\
&&+\frac{6 \pi  \epsilon  a_{{3}/{2}} a_{{7}/{2}}}{k_{CS}}+\frac{3 \pi  \epsilon  \psi _{{3}/{2}} \psi _{{3}/{2}}' N^B_3}{4 k_{CS}}\\
&&-\frac{\pi  \epsilon  \psi _2 \psi _2' N^B_3}{12 k_{CS}}-\frac{\pi  N^B_3 \left(12 a_{{3}/{2}} \epsilon ' a_{{3}/{2}}'+3 \epsilon {a_{{3}/{2}}'}^2+6 a_{{3}/{2}}^2 \epsilon ''+4 \epsilon  a_{{3}/{2}} a_{{3}/{2}}''\right)}{12 k_{CS}}\,.
\end{eqnarray*}

\section{Useful formulae for $A(n,n-1)$ algebra}
\label{degrep}

In this appendix, we derive useful formulae to compute the conformal weight of the degenerate representations of $\cs \cw_n$. For $A(n,n-1)$, we take $\e_{i}$ $(i=1, \ldots, n+1)$ as an orthonormal basis for $\br^{n+1}$ and $\del_{i}$ $(i=1, \ldots, n)$ as that for $\br^{n}$ with the inner product defined as $(\e_{i},\e_{j})=\del_{ij},~(\del_{i},\del_{j})=-\del_{ij},~(\e_{i},\del_{j})=(\del_{i},\e_{j}) = 0$. In this parametrization, there is a redundancy coming from the supertrace condition of $sl(m, n)$, and it results in the following identification \cite{hkv}
\begin{equation}\label{redun}
\sum\limits_{i=1}^{n+1}l_{i}\e_{i}+\sum\limits_{i=1}^{n}k_{i}\del_{i} \sim \sum\limits_{i=1}^{n+1}(l_{i}+t)\e_{i}+\sum\limits_{i=1}^{n}(k_{i}-t)\del_{i}\,,\qquad\forall\, t\,.
\end{equation}
The weight on the left hand side of this identification is denoted as $(l_{1},\ldots, l_{n+1}|k_{1},\ldots, k_{n})$ for simplicity.

We use the pure odd simple root system $\a_{1}, \ldots \a_{2n}$ of $A(n,n-1)$, where $\a$s are represented in terms of $\epsilon_i$ and $\delta_i$ as
\begin{equation}
\a_{2i}=\del_{i}-\e_{i+1}, ~\a_{2i-1}=\e_{i}-\del_{i}\,,
\end{equation}
and the fundamental weights are expressed as
\begin{equation}
\L_{2i-1}=\sum\limits_{j=i}^{n}\big(\del_{j}-\e_{j+1}\big), ~\L_{2i}=\sum\limits_{j=1}^{i}\big(\e_{j}-\del_{j}\big)\,.
\end{equation}
It is then easy to verify the  following identity
\begin{equation}
(\L_{i},~\a_{j})=\del_{ij}\;.
\end{equation}

One can express a general integral dominant weight, which is defined as a linear combination of fundamental weights with non-negative integer coefficients, $\L$, $\r$ and $\nu$, in the $\e_i$ and $\del_i$ basis as
\begin{eqnarray}
\L&=& \sum\limits_{i=1}^{n+1}l_{i}\e_{i}+\sum\limits_{i=1}^{n}k_{i}\del_{i}~\equiv~(l|k)\;,\\
\r   &=& \frac{1}{2} \sum\limits_{i=1}^{n+1}(n+2-2i)\e_{i}- \frac{1}{2} \sum\limits_{i=1}^{n}(n+1-2i)\del_{i}\;,\\
\nu &=& -n {\displaystyle \sum^{n+1}_{i=1}} \epsilon_i
+ (n+1) {\displaystyle \sum^{n}_{i=1}} \delta_i\;,
\end{eqnarray}
where $l_i$ and $k_i$ are constrained for $\Lambda$ to be an integral dominant weight. To see the constraints on $l_i$ and $k_i$, note that $\L$ can be expressed using the fundamental weights as $\L= \sum\limits_{i=1}^{2n}q_{i}\L_{i}$, and the relation between the coefficients $(l_i,k_i)$ and $q_i$ are
\begin{eqnarray}
q_{2j}=-k_{j}-l_{j+1}, \qquad q_{2j-1}=k_{j}+l_{j}\;.
\end{eqnarray}
Therefore, both $-k_{j}-l_{j+1}$ and $k_{j}+l_{j}$ should be non-negative integers. For later convenience, we assume that $l_i \ge 0$ and $k_i \le 0$. One can always achieve this conditions by using the identification \eqref{redun}.

We evaluate the following inner products
\begin{equation}\label{AInnerProduct}
(\Lambda, \Lambda), \qquad (\L,\rho), \qquad (\L,\nu)\;.
\end{equation}
$(\L,\L) = \sum_i l_i^2 - \sum_i k_i^2$ and the only information we use in the main text about this inner product is that it is independent of the value of $n$ and $k$. To evaluate the inner products $(\L,\rho)$ and $(\L,\nu)$, it is very useful to consider the Young supertableaux for Lie superalgebra. Many different versions of Young supertableaux are proposed \cite{djyd, bbyd,hhyd,jhkt,hkv}. The proposal \cite{djyd,bbyd} focus on the ``tensor product''  interpretation  of the Young supertabulaux\fn{Besides, representations of $sl(m|n)$ are always infinite dimensional, even for finite $l_i$ and $k_i$.} and are very convenient for branching rule computation. But in our situation, we want to make use of the ``notation of the highest weight'' interpretation of the Young supertabulaux. So we use the proposal \cite{hhyd} and its extension \cite{hkv}.  The correspondence between the two different representations are discussed in \cite{hhyd}.

The proposed Young superdiagram $F^{(l|k)}$ corresponding to $\L=(l|k)$ consists of a covariant part, which is the Young diagram $F^l$ of weight $l_i$, and a contravariant part, which is the pointwise reflection of the Young diagram $F^{-k}$ of weight $-k_i$ \cite{hkv}. $F^l$ has $l_i$ boxes in the $i^{th}$ row, while $F^{-k}$ has $-k_i$ boxes in the $i^{th}$ column. For example, in $A(6,5)$ algebra, the following Young supertableaux
\begin{equation}\label{egysd}
   \young(::::~,:::~~,:~~~~,~~~~~,:::::~~~~,:::::~~,:::::~)
\end{equation}
corresponds to $\L=(4,2,1,0,0,0,0|0,-1,-2,-2,-3,-4)$.

We evaluate the inner products $(\L,\rho)$ and $(\L,\nu)$, and re-express them in terms of quantities associated with the Young supertableaux. The $(\L,\rho)$ is evaluated as
\begin{eqnarray}
    (\L,\rho) &=&    \frac{1}{2} \sum\limits_{i=1}^{n+1}n l_{i}+ \frac{1}{2} \sum\limits_{i=1}^{n}(n+1)k_{i}+\sum\limits_{i=1}^{n+1}l_{i}-\sum\limits_{i=1}^{n+1}i l_{i}+\sum\limits_{i=1}^{n}i (-k_{i})\nonumber\\
    &=&\frac{1}{2}n B-\frac{1}{2}(n+1)\bar{B}+B-\frac{1}{2}(\sum\limits_{i}c_i^2+B)
   +\frac{1}{2}(\sum\limits_{i}\bar{r}_i^2+\bar{B})\nonumber\\
   &=&\frac{1}{2}n B-\frac{1}{2}(n+1)\bar{B}+\frac{1}{2}(B+\bar{B})+\frac{1}{2}\sum\limits_{i}\bar{r}_i^2
   -\frac{1}{2}\sum\limits_{i}c_i^2\nonumber\\
   &=&\frac{1}{2}n B-\frac{1}{2}(n+1)\bar{B}+\frac{1}{2}\cb+\frac{1}{2}\sum\limits_{i}\bar{r}_i^2
   -\frac{1}{2}\sum\limits_{i}c_i^2\label{ynglr}
\end{eqnarray}
where the $B,\bar{B},\cb$ are numbers of boxes of $F^l,F^{-k},F^{(l|k)}$, respectively, and $c_i, \bar{r}_j$ are number of boxes in column $i$ of $F^l$ and number of boxes in row $j$ of $F^{-k}$. Note that
\begin{equation}\label{BminusbarB}
B - B' = {\displaystyle \sum_{i=1}^{n+1}} l_i + {\displaystyle \sum_{i=1}^{n}} k_i \ge 0
\end{equation}
because of the conditions $l_i \ge 0$ and $k_i+l_i \ge 0$ for any $i$. The $(\L,\nu)$ is evaluated as
\begin{eqnarray}
(\L,\nu) &=& -n {\displaystyle \sum_{i=1}^{n+1}} l_i - (n+1) {\displaystyle \sum_{i=1}^{n}} k_i\nonumber\\
&=& -n B + (n+1) \bar{B}\label{yngln}
\end{eqnarray}
We utilize \eqref{ynglr}, \eqref{yngln} and the fact that $(\Lambda,\Lambda)$ is independent of $n$ and $k$ when evaluating the conformal weights and $U(1)_R$ charge of the degenerate representations.


\begin{thebibliography}{1}

\bibitem{mw}
  A.~Maloney and E.~Witten,
  ``Quantum Gravity Partition Functions in Three Dimensions,''
  JHEP {\bf 1002}, 029 (2010)
  [arXiv:0712.0155 [hep-th]].

\bibitem{gg}
  M.~R.~Gaberdiel and R.~Gopakumar,
  ``An AdS$_3$ Dual for Minimal Model CFTs,''  Phys.\ Rev.\ D {\bf 83}, 066007 (2011)  [arXiv:1011.2986 [hep-th]].  


\bibitem{va}
M.~A.~Vasiliev,
  ``Consistent equation for interacting gauge fields of all spins in (3+1)-dimensions,''
  Phys.\ Lett.\ B {\bf 243} (1990) 378,
  ``Nonlinear equations for symmetric massless higherspin fields in (A)dS(d),''
  Phys.\ Lett.\  B {\bf 567} (2003) 139
  [arXiv:hep-th/0304049].


\bibitem{kp}
  I.~R.~Klebanov and A.~M.~Polyakov,
  ``AdS dual of the critical O(N) vector model,''
  Phys.\ Lett.\ B {\bf 550}, 213 (2002)
  [hep-th/0210114].

\bibitem{hr}
  M.~Henneaux, S.~-J.~Rey,
  ``Nonlinear $W_{infinity}$ as Asymptotic Symmetry of Three-Dimensional Higher Spin Anti-de Sitter Gravity,''
  JHEP {\bf 1012}, 007 (2010).
  [arXiv:1008.4579 [hep-th]].

\bibitem{cfpt}
  A.~Campoleoni, S.~Fredenhagen, S.~Pfenninger, S.~Theisen,
  ``Asymptotic symmetries of three-dimensional gravity coupled to higher-spin fields,''
  JHEP {\bf 1011 } (2010)  007.
  [arXiv:1008.4744 [hep-th]].

\bibitem{gh}
  M.~R.~Gaberdiel, T.~Hartman,
  ``Symmetries of Holographic Minimal Models,''
  JHEP {\bf 1105}, 031 (2011).
  [arXiv:1101.2910 [hep-th]].

\bibitem{cfp}
  A.~Campoleoni, S.~Fredenhagen and S.~Pfenninger,
  ``Asymptotic W-symmetries in three-dimensional higher-spin gauge theories,''
  JHEP {\bf 1109}, 113 (2011)
  [arXiv:1107.0290 [hep-th]].

\bibitem{gghr}
  M.~R.~Gaberdiel, R.~Gopakumar, T.~Hartman and S.~Raju,
  ``Partition Functions of Holographic Minimal Models,''  JHEP {\bf 1108}, 077 (2011)  [arXiv:1106.1897 [hep-th]].  



\bibitem{pr}
  K.~Papadodimas and S.~Raju,
  ``Correlation Functions in Holographic Minimal Models,''
  Nucl.\ Phys.\ B {\bf 856}, 607 (2012)
  [arXiv:1108.3077 [hep-th]].

\bibitem{akp}
  M.~Ammon, P.~Kraus and E.~Perlmutter,
  ``Scalar fields and three-point functions in D=3 higher spin gravity,''
  arXiv:1111.3926 [hep-th].


\bibitem{clm}
  A.~Castro, A.~Lepage-Jutier and A.~Maloney,
  ``Higher Spin Theories in AdS$_3$ and a Gravitational Exclusion Principle,''
  JHEP {\bf 1101}, 142 (2011)
  [arXiv:1012.0598 [hep-th]].

\bibitem{gk}
  M.~Gutperle and P.~Kraus,
  ``Higher spin Black Holes,''
  JHEP {\bf 1105}, 022 (2011)
  [arXiv:1103.4304 [hep-th]].

\bibitem{agkp}
  M.~Ammon, M.~Gutperle, P.~Kraus and E.~Perlmutter,
  ``Spacetime Geometry in Higher Spin Gravity,''
  JHEP {\bf 1110}, 053 (2011)
  [arXiv:1106.4788 [hep-th]].


\bibitem{cggr}
  A.~Castro, R.~Gopakumar, M.~Gutperle and J.~Raeymaekers,
  ``Conical Defects in higher spin Theories,''
  arXiv:1111.3381 [hep-th].

\bibitem{cy}
  C.~-M.~Chang and X.~Yin,
  ``Higher-spin Gravity with Matter in AdS$_3$ and Its CFT Dual,''
  arXiv:1106.2580 [hep-th].


\bibitem{ca}
  C.~Ahn,
  ``The Large N 't Hooft Limit of Coset Minimal Models,''
  JHEP {\bf 1110}, 125 (2011)
  [arXiv:1106.0351 [hep-th]].

\bibitem{gv}
  M.~R.Gaberdiel and C.~Vollenweider,
  ``Minimal Model Holography for SO(2N),''
  JHEP {\bf 1108}, 104 (2011)
  [arXiv:1106.2634 [hep-th]].


\bibitem{Gaberdiel:2012yb}
  M.~R.~Gaberdiel, T.~Hartman and K.~Jin,
  ``Higher Spin Black Holes from CFT,''
  arXiv:1203.0015 [hep-th].


\bibitem{chr}
  T.~Creutzig, Y.~Hikida and P.~B.~Ronne,
  ``Higher spin AdS$_3$ supergravity and its dual CFT,''
  arXiv:1111.2139 [hep-th].

\bibitem{ks}
  Y.~Kazama and H.~Suzuki,
 ``New N=2 Superconformal Field Theories and Superstring Compactification,''
  Nucl.\ Phys.\ B {\bf 321}, 232 (1989).


\bibitem{bdv1}
  E.~Bergshoeff, M.~A.~Vasiliev and B.~de Wit,
  ``The SuperW(infinity) (lambda) algebra,''
  Phys.\ Lett.\ B {\bf 256}, 199 (1991).

\bibitem{bdv2}
  E.~Bergshoeff, B.~de Wit and M.~A.~Vasiliev,
  ``The Structure of the superW(infinity) (lambda) algebra,''
  Nucl.\ Phys.\ B {\bf 366}, 315 (1991).

\bibitem{superW}
  M.~Henneaux, G.~L.~Gomez, J.~Park and S.~-J.~Rey,
  ``Super-W(infinity) Asymptotic Symmetry of Higher-Spin AdS(3) Supergravity,''  arXiv:1203.5152 [hep-th].  

\bibitem{at}
  A.~Achucarro, P.~K.~Townsend,
  ``A Chern-Simons Action for Three-Dimensional anti-De Sitter Supergravity Theories,''
  Phys.\ Lett.\  {\bf B180}, 89 (1986).

\bibitem{w88}
  E.~Witten,
  ``(2+1)-Dimensional Gravity as an Exactly Soluble System,''
  Nucl.\ Phys.\  {\bf B311}, 46 (1988).


\bibitem{hms}
  M.~Henneaux, L.~Maoz, A.~Schwimmer,
  ``Asymptotic dynamics and asymptotic symmetries of three-dimensional extended AdS supergravity,''
  Annals Phys.\  {\bf 282}, 31-66 (2000).
  [hep-th/9910013].

\bibitem{Fradkin:1987ah}
  E.~S.~Fradkin, M.~A.~Vasiliev,
  ``Superalgebra Of Higher-spins And Auxiliary Fields,''
  Int.\ J.\ Mod.\ Phys.\  {\bf A3}, 2983 (1988).

\bibitem{Blencowe:1988gj}
  M.~P.~Blencowe,
  ``A Consistent Interacting Massless Higher Spin Field Theory In D = (2+1),''
  Class.\ Quant.\ Grav.\  {\bf 6}, 443 (1989).


\bibitem{rt}
  T.~Regge, C.~Teitelboim,
 ``Role of Surface Integrals in the Hamiltonian Formulation of General Relativity,''
  Annals Phys.\  {\bf 88}, 286 (1974).

\bibitem{Benguria:1976in}
  R.~Benguria, P.~Cordero and C.~Teitelboim,
  ``Aspects of the Hamiltonian Dynamics of Interacting Gravitational Gauge and Higgs Fields with Applications to Spherical Symmetry,''
  Nucl.\ Phys.\ B {\bf 122}, 61 (1977).



\bibitem{bh}
  J.~D.~Brown, M.~Henneaux,
  ``Central Charges in the Canonical Realization of Asymptotic Symmetries: An Example from Three-Dimensional Gravity,''
  Commun.\ Math.\ Phys.\  {\bf 104}, 207-226 (1986).

\bibitem{b95}
  M.~Banados,
 ``Global charges in Chern-Simons field theory and the (2+1) black hole,''
  Phys.\ Rev.\  {\bf D52}, 5816 (1996).
  [hep-th/9405171].


\bibitem{b98}
  M.~Banados, K.~Bautier, O.~Coussaert, M.~Henneaux, M.~Ortiz,
  ``Anti-de Sitter / CFT correspondence in three-dimensional supergravity,''
  Phys.\ Rev.\  {\bf D58}, 085020 (1998).
  [hep-th/9805165].

\bibitem{b99}
  M.~Banados,
  ``Three-dimensional quantum geometry and black holes,''
  [hep-th/9901148].

\bibitem{Bowcock:1991zk}
  P.~Bowcock and G.~M.~T.~Watts,
  ``On the classification of quantum W algebras,''
  Nucl.\ Phys.\ B {\bf 379}, 63 (1992)
  [hep-th/9111062].

\bibitem{Kac:1977em}
  V.~G.~Kac,
  ``Lie Superalgebras,''
  Adv.\ Math.\  {\bf 26}, 8 (1977).

\bibitem{Dobrev:1985qz}
  V.~K.~Dobrev and V.~B.~Petkova,
  ``Group Theoretical Approach To Extended Conformal Supersymmetry: Function
  Fortsch.\ Phys.\  {\bf 35}, 537 (1987).


\bibitem{itoqr}
  K.~Ito,
  ``Quantum Hamiltonian reduction and N=2 coset models,''  Phys.\ Lett.\ B {\bf 259}, 73 (1991).  


\bibitem{itocp}
  K.~Ito,
  ``N=2 superconformal CP(n) model,''  Nucl.\ Phys.\ B {\bf 370}, 123 (1992).  

\bibitem{itoffr}
  K.~Ito,
  ``Free field realization of N=2 superW(3) algebra,''  Phys.\ Lett.\ B {\bf 304}, 271 (1993)  [hep-th/9302039].  

\bibitem{bs}
  P.~Bouwknegt and K.~Schoutens,
  ``W symmetry in conformal field theory,''  Phys.\ Rept.\  {\bf 223}, 183 (1993)  [hep-th/9210010].  

\bibitem{fl}
  E.~S.~Fradkin and V.~Y.~.Linetsky,
  ``Supersymmetric Racah basis, family of infinite dimensional superalgebras, $SU(\infty + 1|\infty)$ and related 2-D models,''
  Mod.\ Phys.\ Lett.\ A {\bf 6}, 617 (1991).

\bibitem{pc}
Private communications with Constantin Candu and Katsushi Ito.

\bibitem{Gepner:1988wi}
  D.~Gepner,
  ``Scalar Field Theory And String Compactification,''
  Nucl.\ Phys.\ B {\bf 322}, 65 (1989).

\bibitem{Candu:2012jq}
  C.~Candu and M.~R.~Gaberdiel,
  ``Supersymmetric holography on AdS3,''
  arXiv:1203.1939 [hep-th].

\bibitem{djyd}
  P.~H.~Dondi and P.~D.~Jarvis,
  ``Diagram And Superfield Techniques In The Classical Superalgebras,''
  J.\ Phys.\ A A {\bf 14}, 547 (1981).

\bibitem{bbyd}
  A.~Baha Balantekin and I.~Bars,
  ``Representations of Supergroups,''
  J.\ Math.\ Phys.\  {\bf 22}, 1810 (1981).
  A.~Baha Balantekin and I.~Bars,
  ``Branching Rules For The Supergroup $Su(n/m)$ From Those Of $Su(n+m)$,''
  J.\ Math.\ Phys.\  {\bf 23}, 1239 (1982).



\bibitem{hhyd}
  J.~P.~Hurni,
  ``Young Supertableaux Of The Basic Lie Superalgebras,''
  J.\ Phys.\ A A {\bf 20}, 5755 (1987).

\bibitem{jhkt}
    J.~Van der Jeugt, J.~W.~B.~Hughes, R.~C.~King and J.~Thierry-Mieg,
  ``Character Formulae For Irreducible Modules Of The Lie Superalgebras $Sl(m/n)$,''
  J.\ Math.\ Phys.\  {\bf 31}, 2278 (1990).

\bibitem{hkv}
  J.~W.~B.~Hughes, R.~C.~King and J.~Van der Jeugt,
  ``On the Composition factors of Kac modules for the Lie superalgebras sl(m/n),''
  J.\ Math.\ Phys.\  {\bf 33}, 470 (1992).



\end{thebibliography}
\end{document}